\documentclass[twocolumn,aps,prc,superscriptaddress,floatfix]{revtex4}

\usepackage{graphicx}% Include Fig.\ files
\usepackage{longtable}
\usepackage{sidecap}
\usepackage[usenames,dvipsnames]{color}
\usepackage{morefloats}
\usepackage{color}
\usepackage{hypernat}    % for hyperlinking
\usepackage{hyperref}    % for hyperlinking
\usepackage{amsmath}
\usepackage{threeparttablex}

\begin{document}

\preprint{APS/123-QED}

\title{Probing astrophysically important states in $^{26}$Mg nucleus to study neutron sources for the $s$-Process }


\author{R.~Talwar} \email[Electronic Address: ]{rtalwar@anl.gov}
\affiliation{Department of Physics, University of Notre Dame, Notre Dame, Indiana 46556 USA}
\author{T. ~Adachi}
\affiliation{Research Center for Nuclear Physics, Osaka University, Ibaraki, Osaka 567-0047 Japan}
\author{G.~P.~A.~Berg}
\affiliation{Department of Physics, University of Notre Dame, Notre Dame, Indiana 46556 USA}
\author{L.~Bin}
\affiliation{Department of Physics, Osaka University, Toyonaka, Osaka 560-0043 Japan}
\author{S.~Bisterzo}
\affiliation{Department of Physics, University of Turin, Italy}
\affiliation{INAF - Astrophysical Observatory of Turin, Italy}
\author{M.~Couder}
\affiliation{Department of Physics, University of Notre Dame, Notre Dame, Indiana 46556 USA}
\author{R.~J.~deBoer}
\affiliation{Department of Physics, University of Notre Dame, Notre Dame, Indiana 46556 USA}
\author{X.~Fang}
\affiliation{Department of Physics, University of Notre Dame, Notre Dame, Indiana 46556 USA}
\author{H.~Fujita}
\affiliation{Research Center for Nuclear Physics, Osaka University, Ibaraki, Osaka 567-0047 Japan}
\affiliation{Department of Physics, Osaka University, Toyonaka, Osaka 560-0043 Japan}
\author{Y.~Fujita}
\affiliation{Research Center for Nuclear Physics, Osaka University, Ibaraki, Osaka 567-0047 Japan}
\affiliation{Department of Physics, Osaka University, Toyonaka, Osaka 560-0043 Japan}
\author{J.~G\"{o}rres}
\affiliation{Department of Physics, University of Notre Dame, Notre Dame, Indiana 46556 USA}
\author{K.~Hatanaka}
\affiliation{Research Center for Nuclear Physics, Osaka University, Ibaraki, Osaka 567-0047 Japan}
\author{T.~Itoh}
\affiliation{Department of Physics, Osaka University, Toyonaka, Osaka 560-0043 Japan}
\author{T.~Kadoya}
\affiliation{Department of Physics, Kyoto University, Sakyo-ku, Kyoto 606-8501 Japan}
\author{A.~Long}
\affiliation{Department of Physics, University of Notre Dame, Notre Dame, Indiana 46556 USA}
\author{K.~Miki}
\affiliation{Research Center for Nuclear Physics, Osaka University, Ibaraki, Osaka 567-0047 Japan}
\author{D.~Patel}
\affiliation{Department of Physics, University of Notre Dame, Notre Dame, Indiana 46556 USA}
\author{M.~Pignatari}
\affiliation{Konkoly Observatory, Research Center for Astronomy and Earth Sciences, Hungarian Academy of Sciences}
\author{Y. ~Shimbara}
\affiliation{CYRIC, Tohoku University, Aramaki, Aoba, Sendai, 980-8578 Japan}
\author{A.~Tamii}
\affiliation{Research Center for Nuclear Physics, Osaka University, Ibaraki, Osaka 567-0047 Japan}
\author{M.~Wiescher}
\affiliation{Department of Physics, University of Notre Dame, Notre Dame, Indiana 46556 USA}
\author{T.~Yamamoto}
\affiliation{Research Center for Nuclear Physics, Osaka University, Ibaraki, Osaka 567-0047 Japan}
\author{M.~Yosoi}
\affiliation{Research Center for Nuclear Physics, Osaka University, Ibaraki, Osaka 567-0047 Japan}
%\thanks{}


\date{\today}

\begin{abstract}

\begin{description}

\item[Background]: The $^{22}$Ne($\alpha$,n)$^{25}$Mg reaction is  the dominant neutron source for the slow neutron capture process ($s$-process) in massive stars and contributes, together with the $^{13}$C($\alpha$,n)$^{16}$O, to the production of neutrons for the $s$-process in Asymptotic Giant Branch (AGB) stars. However, the reaction is endothermic and competes directly with the $^{22}$Ne($\alpha,\gamma)^{26}$Mg radiative capture. The uncertainties for both reactions are large owing to the uncertainty in the level structure of $^{26}$Mg near the alpha and neutron separation energies. These uncertainties are affecting the s-process nucleosynthesis calculations in theoretical stellar models.     
\item[Purpose]: Indirect studies in the past have been successful in determining the energies, $\gamma$-ray and neutron widths of the $^{26}$Mg states in the energy region of interest. But, the high Coulomb barrier hinders a direct measurement of the resonance strengths, which are determined by the $\alpha$-widths for these states. The goal of the present experiments is to identify the critical resonance states and to precisely measure the $\alpha$-widths by $\alpha$ transfer techniques .  
\item[Methods]: The $\alpha$-inelastic scattering and $\alpha$-transfer measurements were performed on a solid $^{26}$Mg target and a $^{22}$Ne gas target, respectively, using the Grand Raiden Spectrometer at the Research Center for Nuclear Physics in Osaka, Japan. The ($\alpha$,$\alpha$') measurements were performed at 0.45$^{\circ}$, 4.1$^{\circ}$, 8.6$^{\circ}$ and 11.1$^{\circ}$ and the ($^6$Li,$d$) measurements at 0$^{\circ}$ and 10$^{\circ}$. The scattered $\alpha$ particles and deuteron were detected by the focal plane detection system consisting of multi-wire drift chambers and plastic scintillators. The focal plane energy calibration allowed the study of $^{26}$Mg levels from E$_x$ = 7.69-12.06 MeV in the ($\alpha,\alpha'$) measurement and E$_x$ = 7.36-11.32 MeV in the ($^6$Li,$d$) measurement.
\item[Results]: Six levels (E$_x$ = 10717 keV , 10822 keV, 10951 keV, 11085 keV, 11167 keV and 11317 keV)  were observed above the $\alpha$-threshold in the region of interest (10.61 - 11.32 MeV).  The alpha-widths were calculated for these states from the experimental data. The results were used to determine the $\alpha$-capture induced reaction rates.
\item[Conclusion]: The energy range above the $\alpha$ threshold in $^{26}$Mg was investigated using a high resolution sprectrometer. A number of states were observed for the first time in $\alpha$ scattering and $\alpha$ transfer reactions. The excitation energies and spin-parities were determined. Good agreement is observed for previously known levels in $^{26}$Mg. From the observed resonance levels the E$_x$ = 10717 keV state has a negligible contribution to the $\alpha$-induced reaction rates. The rates are dominated in both reaction channels by the resonance contributions of the states at E$_x$ = 10951, 11167 and 11317 keV. The E$_x$ =11167 keV has the most appreciable impact on the ($\alpha,\gamma$) rate and therefore plays an important role for the prediction of the neutron production in s-process environments. 
\end{description}

\end{abstract}

\pacs{Valid PACS appear here}% PACS, the Physics and Astronomy

\maketitle


\section{INTRODUCTION} \label{sec:intro}

The $^{22}$Ne($\alpha,n)^{25}$Mg is one of the dominant neutron sources for the $s$-process in stars ~\citep{RevModPhys.83.157}. The reaction occurs in He-burning environments in massive stars (M $\textgreater$ 8M$_{\odot}$) and in low- and intermediate-mass stars during asymptotic Giant Branch phase. During He burning the bulk of $^{22}$Ne is made by the reaction sequence $^{14}$N($\alpha,\gamma)^{18}$F($\beta^+,\nu)^{18}$O($\alpha,\gamma)^{22}$Ne. This sequence is initiated on the high abundance of the nucleus $^{14}$N in the ashes of the CNO cycle during the preceding hydrogen burning phase of main sequence stars ~\citep{Rai91}, ~\citep{Gal98}. 

The understanding of $s$-process nucleosynthesis is of considerable importance. The $s$-process is responsible for the formation of about half of the elements heavier than iron ~\citep{doi:10.1146/annurev.astro.46.060407.145207}. It proceeds along the line of stability via a sequence of neutron capture reactions on stellar seed material followed by the $\beta$-decay of short-lived reaction products. 

It determines, together with the rapid neutron capture process ($r$-process) ~\citep{Thielemann2011346}, the distribution of most of the elements heavier than Fe in the solar system. The dominant astrophysical source of the $r$-process is still matter of debate ~\citep{vandeVoort11022015}, ~\citep{Wehmeyer11092015}, ~\citep{Ces15}, and the large nuclear physics uncertainties affecting the $r$-process path are limiting the predictive power of theoretical $r$-process predictions. 

The residual method is a critical tool for extracting the $r$-process pattern in the solar system, which is given by the solar abundances aftre removing the $s$-process contribution ~\citep{0004-637X-525-2-886}, ~\citep{MNR:MNR19484}. In general, the $r$-process residual identified in the solar system has been shown to be compatible with the $r$-process abundance pattern observed in very old metal poor stars ~\citep{doi:10.1146/annurev.astro.46.060407.145207}, keeping into account a number of relevant differences ~\citep{0004-637X-601-2-864}, ~\citep{0004-637X-724-2-975}, ~\citep{0004-637X-787-1-10}. The detailed understanding of the of $s$-process abundance distribution is therefore critical for a reliable identification of all possible contributions responsible for these deviations.

At low metallicity the elemnetal products of the $s$-process nucleosynthesis in AGB stars can be directly observed in carbon-enhanced metal poor stars ~\citep{doi:10.1146/annurev.astro.46.060407.145207}, ~\citep{Bisterzo01052012}, ~\citep{MAPS:MAPS1411}, ~\citep{2015arXiv150301990H}, in post-AGB stars ~\citep{Sme12}, ~\citep{Sme14} and in Ba stars ~\citep{0004-637X-557-2-802}. For several cases the observations seem to agree well with theoretical model predictions ~\citep{Bus99}, ~\citep{PAS:9305903}, while for other cases there are problems to reproduce the observations ~\citep{Van03}, ~\citep{refId01}. At low metallicity it might be possible to observe the $s$-process activated in fast rotating massive stars ~\citep{1538-4357-687-2-L95}, ~\citep{Ces13}, ~\citep{Fri12}. At metallicities much closer to solar, the chemical composition of planetary nebulae is affected by $s$-process nucleosynthesis  in the central AGB star \citep{0004-637X-659-2-1265}. Of great importance is the observation and measurement of isotopic abundances of $s$ process products that can be directly derived from the analysis of meteoritic inclusions \citep{doi:10.1146/annurev.earth.26.1.147}

The $s$-process distribution in the solar system has been divided in three components: between Fe and Sr there is the weak s-process component, associated to the s-process production in massive stars ~\citep{Rai91}, ~\citep{0004-637X-655-2-1058}, ~\citep{Pig10}. Between the Sr neutron-magic peak and Pb there is the the main s-process component, and half of the solar Pb208 is made by the strong s-process component ~\citep{Gal98}. Both the strong and the main component are made in AGB stars, but at different metallicities.

In low mass (1.5-3 M$_{\odot}$) AGB stars, $^{13}$C($\alpha$,n)$^{16}$O is the main neutron source during the inter pulse period, while $^{22}$Ne($\alpha$,n)$^{25}$Mg is marginally activated during advanced thermal pulses ($T$ $\approx$ 0.3 GK) ~\citep{Gal98}. In case of AGB stars with intermediate initial mass (M $\textgreater$ 3 M$_{\odot}$), much higher temperatures are readily achieved ($T$ $\approx$ 0.35 GK) thereby efficiently activating the $^{22}$Ne($\alpha$,n)$^{25}$Mg reaction ~\citep{PAS:9305903}, ~\citep{0004-637X-785-1-77}. The $^{13}$C($\alpha$,n)$^{16}$O plays a marginal role in this AGB mass range ~\citep{0004-637X-785-1-77}.

The dominant site for the weak $s$-process component is the core-helium burning in massive stars. The neutron flux is expected to be much lower than in AGB stars, therefore, only light $s$-process isotopes (A $\textless$ 90) are generated during this phase. The $^{22}$Ne($\alpha$,n)$^{25}$Mg reaction is considered the most important neutron source ~\citep{0004-637X-655-2-1058}. However, due to the negative Q-value (Q=-0.478 MeV) of the reaction, higher temperatures are required to warrant a sufficiently high neutron flux. Therefore the main neutron production is expected towards the final phase of core helium burning when the helium fuel has substantially declined and the core has started to contract under its own gravitational weight. This contraction increases the temperature and density conditions and turns the  $^{22}$Ne($\alpha$,n)$^{25}$Mg reaction into a viable neutron source.  

Because of the rapid decline in helium fuel, not all $^{22}$Ne might be consumed ~\citep{Pra90}. Therefore the $\alpha$ particles generated via $^{12}$C($^{12}$C,$\alpha)^{20}$Ne reaction channel during the subsequent C-burning phase will re-activate the $^{22}$Ne($\alpha$,n)$^{25}$Mg reaction ~\citep{Arn69}. Along with $\alpha$-particles, protons also become readily available at the same time via the $^{12}$C($^{12}$C,p)$^{23}$Na reaction. Hence, in this scenario, the $^{22}$Ne(p,$\gamma)^{23}$Na becomes the main competitor of the $^{22}$Ne neutron source ~\citep{Pig10}. Nonetheless, the $s$-process nucleosynthesis occurs during convective shell C-burning at a high neutron density and with neutron exposure comparable to that in the previous He-core burning stage ~\citep{Pig10}, ~\citep{Rai91}.

A recent paper by Liu et al. \citep{Lui} analyzed the strength of the $^{22}$Ne($\alpha$,n)$^{25}$Mg neutron source on the basis of the observed barium isotopic abundance distribution in meteoritic inclusions. They found that the reaction rate is most likely smaller than predicted in the NACRE reaction rate tabulation ~\citep{Angulo19993} that was based on an earlier analysis of the reaction rate \citep{1994ApJ...437..396K}. This conclusion is based on the lower neutron flux conditions required for matching the observed barium isotope abundances. This is an interesting assessment but there may be other explanations that are associated with the complex nature of the $^{22}$Ne($\alpha$,n)$^{25}$Mg neutron source. 

As already pointed out in earlier work \citep{1994ApJ...437..396K}, an important aspect in the discussion of the $^{22}$Ne($\alpha$,n)$^{25}$Mg reaction as an effective neutron source is the competing $^{22}$Ne($\alpha,\gamma$)$^{26}$Mg radiative capture process. Radiative capture reactions are facilitated through the electro-magnetic forces and are therefore typically weaker than nuclear reactions with cross sections based on the strong force. However, the $^{22}$Ne($\alpha,\gamma$)$^{26}$Mg reaction has a positive Q-value and therefore is effective during the entire helium burning phase where it can substantially reduce the amount of $^{22}$Ne before the $^{22}$Ne($\alpha$,n)$^{25}$Mg reaction with its negative $Q$-value will start operating. This may not affect the neutron production during the rapidly occuring helium flashes in TP-AGB stars, but it may significantly affect the weak $s$-process nucleosynthesis that operates on much longer time-scales. If the  $^{22}$Ne($\alpha,\gamma$)$^{26}$Mg is sufficiently strong, the limited $^{22}$Ne abundance may be too low for efficient neutron production in the late phase of helium burning and reduce neutron production during carbon burning.  The overall neutron yield is therefore not only governed by the abundance of $^{22}$Ne but also by the branching ratio between the $\gamma$- and $n$-exit channels. For both channels the reaction rates are influenced by the resonance levels in the $^{26}$Mg compound nucleus. A strong $^{22}$Ne($\alpha$,$\gamma$)$^{25}$Mg reaction would reduce the overall $^{22}$Ne abundance during low temperature He burning and reduce the neutron flux at higher temperature conditions. Therefore a complete understanding of both reactions is necessary to understand this interplay between these two reaction channels. The goal of this paper is to deliver a comprehensive study of these levels above the $\alpha$ threshold in $^{26}$Mg and explore the impact on the respective reaction rates.

\section{LEVEL DENSITY AND ALPHA CLUSTER STRUCTURE}\label{sec:nucstruc}

While stellar hydrogen burning is primarily facilitated through direct capture and resonance configurations associated with pronounced single particle structure in the compound nuclei of radiative capture and nuclear reactions in the pp-chains and the CNO cycles, reactions in stellar helium burning are characterized by the contributions of resonances that can be identified as $\alpha$ cluster configurations in the compound nuclei \citep{doi:10.1016/j.ppnp.2004.09.002}. Such $\alpha$-cluster configurations are expected in even-even nuclei near the threshold for break up into an $\alpha$-particle plus the residual core nucleus as expressed by the ' Ikeda rule ' ~\citep{ikeda1968systematic}. An alpha particle represents a cluster of two protons and two neutrons. Such closed shell configuration makes alpha particle particularly stable in self-conjugate nuclei owing to pairing effect. 

The most famous example is the ground state of $^8$Be and the Hoyle-state, a pronounced three $\alpha$ cluster configuration in $^{12}$C that corresponds to a 0$^+$ resonance level at 7.65 MeV. Both of these levels facilitate the triple-alpha process leading to the formation of $^{12}$C in stars \citep{doi:10.1103/PhysRevLett.106.192501}. Other pronounced $\alpha$ cluster resonance configurations have been found in $^{16}$O, influencing the $^{12}$C($\alpha,\gamma$)$^{16}$O reaction \citep{doi:10.1103/PhysRevLett.83.4025} and in $^{22}$Ne responsible for the fast conversion of $^{18}$O via the $^{18}$O($\alpha,\gamma$)$^{22}$Ne reaction ~\citep{Dabab}.     

There are a number of similar cases of low energy resonances with pronounced $\alpha$ cluster structure and indeed, like in $^{22}$Ne, other T=1 (N$\neq$Z) nuclei such as $^{18}$O ~\citep{Joh09}, and $^{26}$Mg ~\citep{Giesen199395} exhibit resonance features that correspond to $\alpha$-cluster states. The identification of $\alpha$-clusters are based on small single particle and large alpha spectroscopic factors. Such levels are characterized by large resonance strength in alpha capture and transfer reactions. However, in low energy radiative capture measurements the strength is suppressed by the Coulomb-barrier, while alpha transfer reactions reflect the full alpha strength distribution.  

Considerable efforts have been made in the past to perform direct measurements of the $^{22}$Ne($\alpha$,n)$^{25}$Mg (~\citep{Ashery1969481},~\citep{PhysRevC.7.2432}, ~\citep{Dro91}, ~\citep{PhysRevC.43.2849}, ~\citep{Wol89} and ~\citep{PhysRevLett.87.202501}) and the $^{22}$Ne($\alpha,\gamma)^{26}$Mg (~\citep{Wol89} and ~\citep{PhysRevC.11.1525}) reactions. Unfortunately, in the astrophysical region of interest, $\alpha$-penetrability is largely suppressed by the Coulomb barrier and very difficult to measure because of the cosmic and beam-induced background in the detector materials. Only upper limits could therefore be obtained for the neutron and $\gamma$ yield at energies below the lowest directly observed resonance at E$_{\alpha}$ = 832 keV (E$_x$ = 11.318 MeV). 

A number of scattering and transfer measurements (~\citep{Moss1976429},~\citep{PhysRevC.39.311}, ~\citep{VanDerBorg1981243}, ~\citep{Giesen199395}, ~\citep{PhysRevC.76.025802}, ~\citep{PhysRevC.85.044615} and ~\citep{PhysRevC.80.055803}) have been performed to investigate the level structure of $^{26}$Mg above the $\alpha$-threshold. The $^{26}$Mg($\alpha,\alpha'$)$^{26}$Mg measurement by Borg et al. ~\citep{VanDerBorg1981243}  exhibited poor resolution ($\sim$ 120 keV) and the $^{22}$Ne($^6$Li,d)$^{26}$Mg measurements by Giesen et al. ~\citep{Giesen199395} and Ugalde et al. ~\citep{PhysRevC.76.025802} were handicapped by high beam induced background causing huge contamination peaks in the astrophysical region of interest. These measurements were complemented by the study of additional reaction channels such as $^{25}$Mg(n,$\gamma$)$^{26}$Mg  ~\citep{PhysRevC.85.044615}, ~\citep{PhysRevC.14.1328}, inelastic proton scattering measurements on $^{26}$Mg ~\citep{Fuj} using the Grand Raiden Spectrometer at RCNP, Osaka, Japan as well as studies of $^{26}$Mg($\gamma,\gamma$')$^{26}$Mg by Longland et al. ~\citep{PhysRevC.80.055803} and deBoer et al. ~\citep{PhysRevC.82.025802} and $^{26}$Mg($\gamma,n$)$^{25}$Mg measurement by deBoer et al. ~\citep{PhysRevC.89.055802}. The results did provide additional information on the n- and $\gamma$- widths of the near threshold levels and added important spin parity information about the alpha-unbound states in $^{26}$Mg. However, the critical parameter that needs to be determined for deriving the $^{22}$Ne +$\alpha$ resonance strengths is the $\alpha$ partial width of these states.

 In the present work, $\alpha$-inelastic scattering (with improved resolution of $\simeq$ 65 keV) and $\alpha$-transfer via ($^6$Li,$d$) (with a well-defined background shape using thick target yield function ~\citep{RevModPhys.20.236}) have been used to probe the $^{26}$Mg nucleus using the Grand Raiden Spectrometer. The main goal is to determine the resonance energies and $\alpha$-widths for levels above the $\alpha$-threshold, serving as input parameters into the $^{22}$Ne+$\alpha$ capture reaction rate calculation. The $\alpha$-widths will also help establish the predicted alpha cluster structure for these levels.\\

\section{EXPERIMENTAL SETUP} \label{sec:exp}

To study the low energy resonances in $^{22}$Ne+$\alpha$, $^{26}$Mg($\alpha,\alpha')^{26}$Mg and $^{22}$Ne($^6$Li,d)$^{26}$Mg reactions have been measured using the high resolution Grand Raiden (GR) spectrometer at the Research Center for Nuclear Physics (RCNP) in Osaka, Japan. Both experiments were designed to cover the energy range of interest ($E_x$ = 10.61 MeV - 11.32 MeV) in the $^{26}$Mg nucleus. 

For the $\alpha$-inelastic scattering measurement, a self-supporting $^{26}$Mg target (enriched to 99.4\%) of thickness 1.16 mg/cm$^2$ was used. Since $^{26}$Mg oxidizes rapidly when exposed to air, impurity peaks corresponding to $^{16}$O were observed in addition to those owing to Carbon contamination. Background runs were taken on CH$_2$ (1.13 mg/cm$^2$) and Mylar ((C$_{10}$H$_8$O$_4$)$_n$) (1 mg/cm$^2$) targets. For focal plane energy calibration, the $^{25}$Mg($\alpha,^3$He)$^{26}$Mg reaction was measured that populated a significant part of the focal plane with well known low energy resonances in $^{26}$Mg ~\citep{PhysRevC.34.1530}.

A 206 MeV $\alpha$-beam was generated using the coupled AVF and Ring cyclotrons and was transported via the fully dispersion matched WS beam line ~\citep{Wakasa200279} to the target chamber upstream of the GR spectrometer. The new WS beam line has been designed to satisfy all the required matching conditions ~\citep{Fujita1997274}: focussing condition, lateral dispersion matching, kinematic correction and angular dispersion matching. For the present measurements, the faint-beam method was applied wherein a low intensity beam (10$^3$ particles/s) was directly sent into the spectrometer, placed at 0$^{\circ}$, so that the matching conditions could be diagnosed using the beam properties in the focal plane ~\citep{Fujita200217}. This technique ensured that the final resolution was not limited by the momentum spread (150 - 200 keV) of the beam exiting from the cyclotron. 

\begin{figure}
 \includegraphics[width= 1.0\columnwidth]{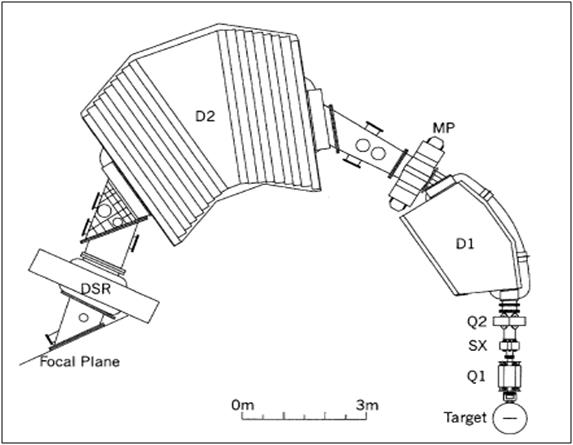}
    \caption[]{Schematic layout of the Grand Raiden Spectrometer at RCNP. The DSR was not used in the present experiments, but it is a part of the permanent installation. Figure from reference ~\citep{Wakasa200279}.}
    \label{fig:grand-raiden}
 \end{figure}

The scattered alpha particles emerging from the target were momentum analyzed by the GR spectrometer (Fig. ~\ref{fig:grand-raiden}) with a high resolving power of p/$\Delta$p = 37000 ~\citep{Fujiwara1999484}. They were detected at the focal plane detection system, which consisted of two multi-wire drift chambers (MWDCs) and a stack of 3 mm and 10 mm thick plastic scintillators (PS1 and PS2) along with a 2 mm thick Aluminium absorber placed between the two scintillators. The MWDCs provided position and angular information in the horizontal and vertical directions and the scintillators gave time of flight and energy loss information for particle identification. In order to precisely reconstruct the vertical component of the scattering angle at and near 0$^{\circ}$, the off-focus mode ~\citep{Fujita200155} was employed. A sieve-slit (multi-hole aperture) was used to perform the angle calibration measurement. A special beam exit pipe was incorporated in the exit window of the focal plane to collect the beam at 0.45$^{\circ}$ in the Faraday cup downstream of the focal plane detector. The  Faraday cup downstream of quadrupole Q1 was used for 2$^{\circ}$-6$^{\circ}$ settings of the spectrometer and for higher angles, the cup inside the scattering chamber was used.

For the $\alpha$-transfer measurement, highly enriched $^{22}$Ne gas (enrichment $\textgreater$ 99\%) was pressurized to 0.2 atm in a gas-cell using a gas handling system ~\citep{Matsubara2012122}. The cell body was machined from copper and the gas was filled into a volume measuring 44 mm by 14 mm by 10 mm. Aramid (C$_{14}$O$_2$N$_2$Cl$_2$H$_8$) films of thickness 4 $\mu$m were used as entrance and exit windows to cover the aperture in the cell body. In addition to the $^{22}$Ne gas target, ($^6$Li,$d$) measurements were also performed on the 4 $\mu$m Aramid foil, $^{16}$O and $^{20}$Ne gas targets to identify background peaks and perform focal plane energy calibration using the well known low energy peaks ~\citep{PhysRevC.24.2556}, ~\citep{Anantaraman1977474}. 

The $^6$Li beam with an energy of $E_{\text{lab}}$ = 82.3 MeV was generated using the AVF cyclotron. All other experimental procedures and set-up were the same as those for the $\alpha$-inelastic scattering measurement. Exceptions were (a) the use of a stack of two plastic scintillators each of thickness 10 mm and (b) the 0$^{\circ}$ Faraday cup was placed inside the first dipole D1 because the B$\rho$ ratio of deuteron to $^6$Li is 1.7.\\

\section{DATA ANALYSIS AND RESULTS} \label{sec:dat}

The 5\% momentum acceptance of the Grand Raiden spectrometer allowed an excitation energy range coverage of 3 - 12 MeV in the ($\alpha,\alpha'$) experiment and 7 - 12 MeV in the ($^6$Li,$d$) experiment, for a single magnetic field setting. 

Appropriate gates were set on the scattered alpha particles and deuteron in the time of flight spectrum as well as the energy loss spectrum coming from the plastic scintillators. This reduced the background coming from multiple scattering events. The first order dependence of the resolution on the energy spread of the incident beam was eliminated using the dispersion matching technique ~\citep{Fujita200217}. However, the effects of reaction kinematics and higher order magnetic aberrations had to be corrected for during the offline analysis. This resulted in a resolution of 65 keV for the ($\alpha,\alpha'$) measurement and 100 keV for the ($^6$Li,$d$) measurement. These values include the effects of energy losses through the solid $^{26}$Mg target (22 keV), the $^{22}$Ne gas target (11 keV), and energy straggling in the entrance and exit foils of the gas cell, along with the effects of angular straggling of the beam through these foils.

\subsection{Energy Calibration and Peak Identification}

\begin{figure*}
\includegraphics[width=1.5\columnwidth]{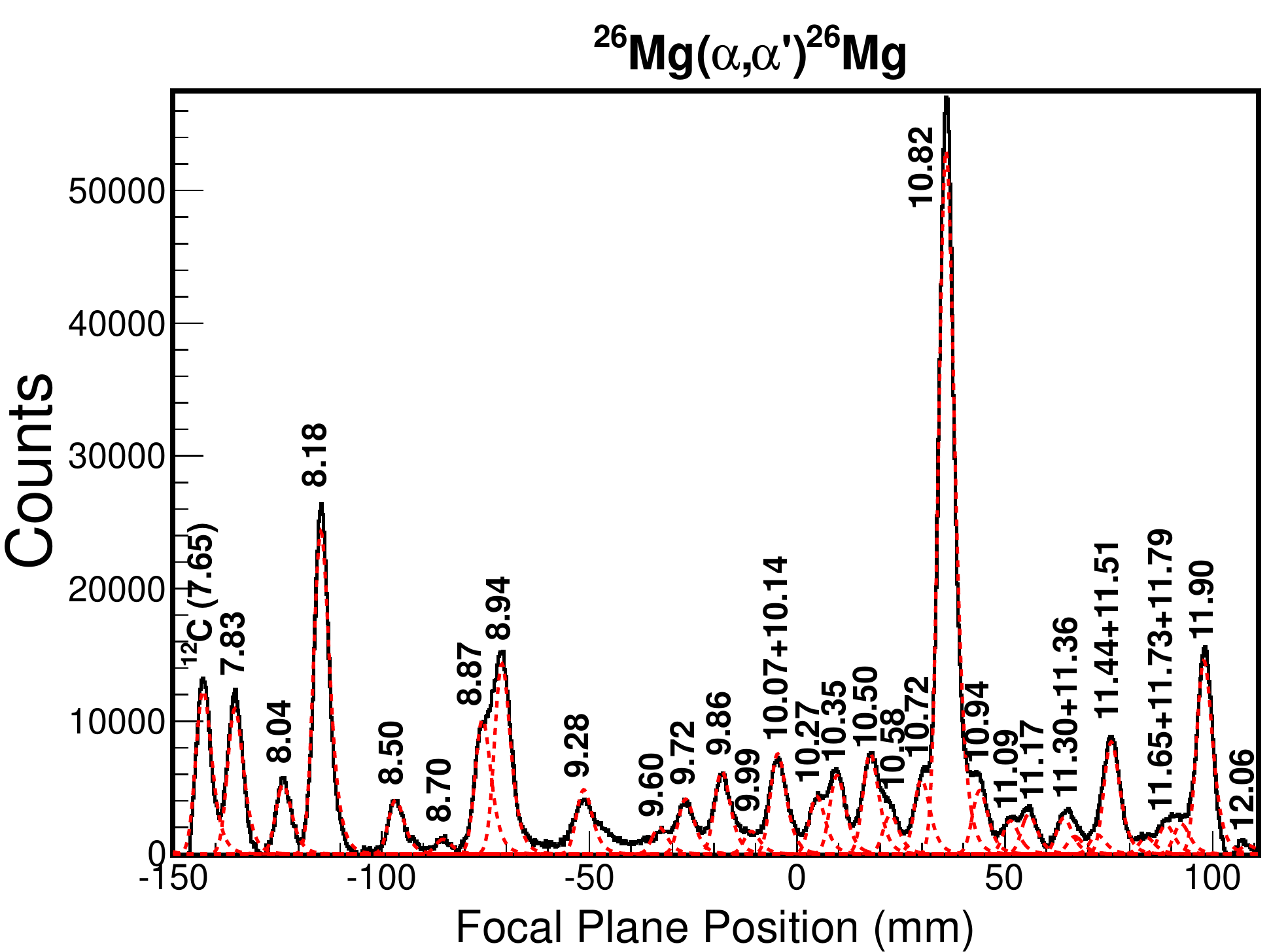}
\caption{(Color online) Background subtracted spectrum showing $^{26}$Mg peaks coming from the  ($\alpha,\alpha'$) measurement at a spectrometer angle of 0.45$^{\circ}$. All energies are in MeV. The red dashed lines represent the individual peaks.}
\label{fig:alpha_peaks}
\end{figure*}

\begin{figure*}
\includegraphics[width=1.5\columnwidth]{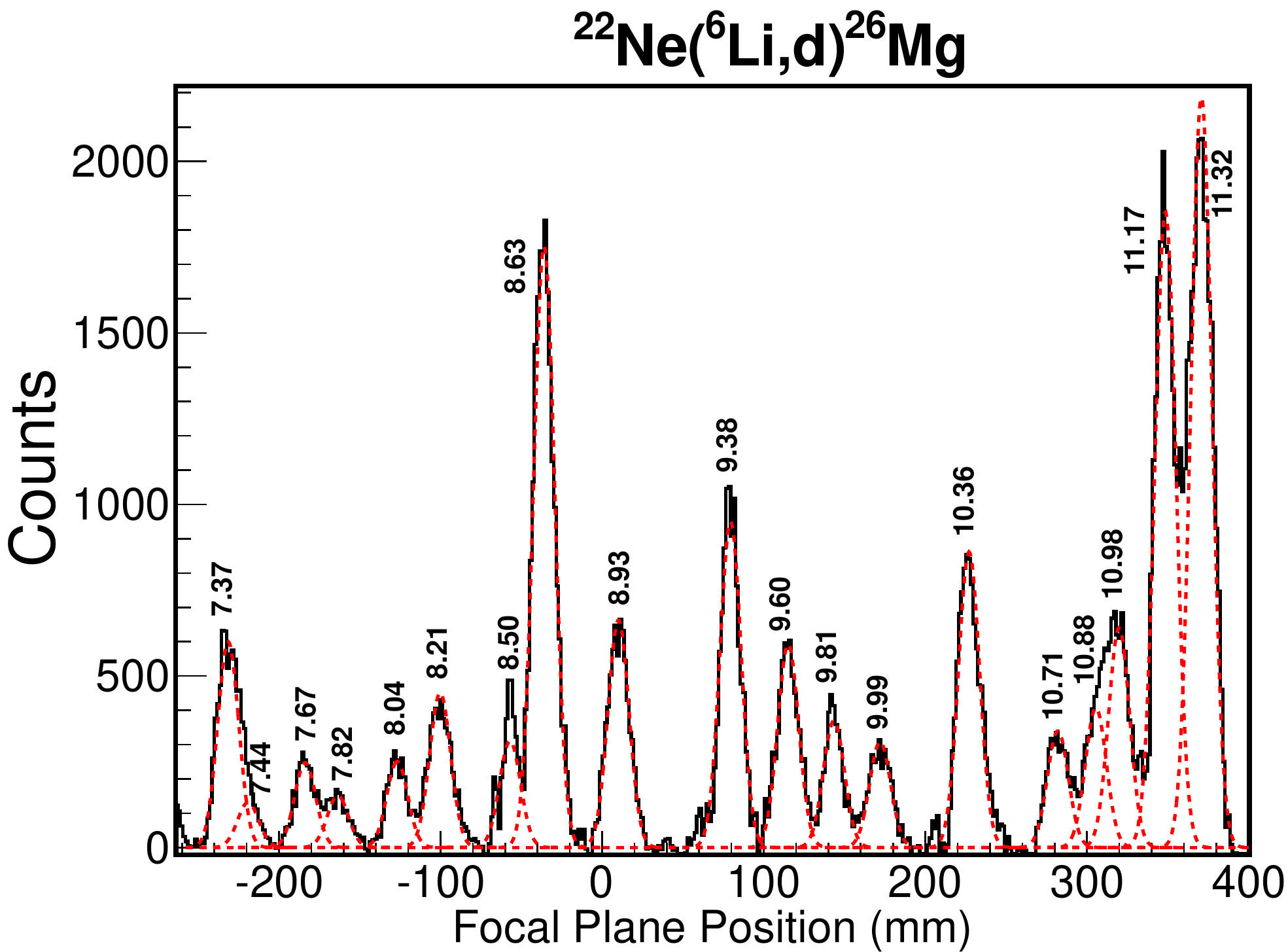}
\caption{(Color online) Background subtracted spectrum showing $^{26}$Mg peaks coming from the 0$^{\circ}$ ($^6$Li,$d$) measurement.   All energies are in MeV. The red dashed lines represent the individual peaks.}
\label{fig:deuteron_peaks}
\end{figure*}

Establishing a well defined relationship between the magnetic rigidity (B$\rho$) of the outgoing particle and its corresponding position at the focal plane was an important prerequisite to accurately determine the excitation energies associated with the inelastic scattering and alpha-transfer peaks.  Precise determination of the focal plane position was achieved using an asymmetric Gaussian function plus polynomial background to fit the ($\alpha,\alpha'$) peaks and a Gaussian function plus arctangent background ~\citep{fowler1948gamma} to fit the ($^6$Li,$d$) peaks and the thick target Aramid background. Magnetic rigidities were determined for the well-known low-lying states populated in $^{26}$Mg ~\citep{Endt19981} via the $^{25}$Mg($\alpha,^3$He) reaction ~\citep{PhysRevC.34.1530}, in $^{20}$Ne ~\citep{Tilley1998249} via the  $^{16}$O($^6$Li,d) reaction ~\citep{PhysRevC.24.2556} and in $^{24}$Mg ~\citep{Firestone20072319} via the $^{20}$Ne($^6$Li,d) reaction ~\citep{Anantaraman1977474}. Using these peaks, mainly linear calibration functions with small quadratic terms were established that allowed identification of $^{26}$Mg peaks ranging from $E_x$ = 7.69-12.06 MeV at 0.45$^{\circ}$, 4.1$^{\circ}$, 8.6$^{\circ}$ and 11.1$^{\circ}$ in the ($\alpha,\alpha'$) measurement (Fig. ~\ref{fig:alpha_peaks}) and $E_x$ = 7.36-11.32 MeV at 0$^{\circ}$ and 10$^{\circ}$ in the ($^6$Li,$d$) measurement (Fig. ~\ref{fig:deuteron_peaks}). The results for the excitation energies were determined by taking a weighted average of the energies measured at different angles. The errors associated with these energies were computed as a quadratic combination of the statistical error (3-8 keV for ($\alpha,\alpha'$) measurement and 12-30 keV for ($^6$Li,$d$) measurement) and the systematic error (5-10 keV for both measurements) arising from uncertainties in energy calibration, energy loss calculations using SRIM ~\citep{SRIM}, target inhomogeneities, reaction angle determinations, and the number of counts in the peak. 
 
The observed excitation energies are presented in three tables: (i) energy levels below the $\alpha$-threshold (10.615 MeV) in Table ~\ref{tab:energy_levels_1} , (ii) energy levels above the $\alpha$-threshold (10.615 MeV) and below the neutron threshold (11.093 MeV)  in Table ~\ref{tab:energy_levels_2} and (iii) energy levels above the neutron threshold (11.093 MeV) in Table ~\ref{tab:energy_levels_3}. In all three tables, the observed levels were compared with previous results.

\begin{table*}
\caption{The $^{26}$Mg excitation energies measured below the $\alpha$-threshold (10.615 MeV) in the present work along with the comparison with the values from the compilation. The numbers in parenthesis are the uncertainties in the last digits of the energy values.}
\label{tab:energy_levels_1}
\begin{ruledtabular}
\begin{tabular}{c c c | c c c }
\multicolumn{2}{c}{Present Work} & Endt98 ~\citep{Endt19981} & \multicolumn{2}{c}{Present Work} & Endt98 ~\citep{Endt19981}\\ [0.2cm]
$^{26}$Mg($\alpha,\alpha')^{26}$Mg & $^{22}$Ne($^6$Li,d)$^{26}$Mg & Compilation & $^{26}$Mg($\alpha,\alpha')^{26}$Mg & $^{22}$Ne($^6$Li,d)$^{26}$Mg & Compilation\\[0.2cm] 
$E_x$(keV) & $E_x$(keV) & $E_x$(keV) & $E_x$(keV) & $E_x$(keV) & $E_x$(keV)\\ 
\hline\hline\\
& & 7200 (20) & & & 9371 (2) \\
& & 7242 (1) & & 9383 (16)& 9383 (1) \\
& & 7261.39 (4) & & & 9427.74 (7) \\
& & 7282.74 (5) & & & 9471 (2) \\
& & 7348.87 (5) & & & 9541 (1)\\
& 7365 (13) & & & 9560 (3)\\
& & 7395 (1) & & & 9574.02 (6)\\
& & 7428 (3) & & & 9579 (3)\\
& & 7541.73 (5) & 9604 (9)& 9595 (32) & 9590 (2)\\
& & 7677 (1) & & & 9617.0 (9)\\
7688 (7) & 7671 (16) & 7697.3 (6) & & & 9681 (2)\\ 
& & 7725.74 (16) & 9718 (7) & & 9714 (3)\\
& & 7773 (1) & & & 9771 (2)\\
& & 7816 (2) & & & 9779 (3)\\
7827 (6) & 7821 (22) & 7824 (3) & & & 9814 (2)\\
& & 7840 (2) & & & 9829 (1)\\
& & 7851 (3) & 9863 (6) & & 9856.52 (6)\\
& & 7953 (1) & & & 9883 (3)\\
8035 (7) & 8040 (13) & 8033 (2) & & & 9902 (2)\\
& & 8052.9 (6) & & & 9927 (2)\\
8185 (9) & 8214 (14) & 8184.96 (10) & & & 9939 (2)\\
& & 8201 (1) & & & 9967 (2)\\
& & 8227.56 (16) & & & 9982 (2)\\
& & 8250.70 (10) & 9993 (9) & 9987 (18) & 9989 (1)\\
& & 8399 (3) & & & 10040 (2)\\
& & 8458.87 (13) & 10067 (7) & & 10069 (2)\\
& & 8464 (2) & & & 10102.41 (15)\\
& & 8472 (1) & & & 10126.70 (10)\\
8497 (8) & & 8503.74 (9) & 10136 (8) & & 10136 (3)\\
& & 8532.27 (9) & & & 10148 (2)\\
& & 8577 (3) & & & 10159 (3)\\
8626 (7) & 8625 (15) & 8625 (1) & & & 10184 (2)\\
& & 8670 (1) & & & 10220.1 (3)\\
8703 (6) & & 8705.73 (9) & & & 10234 (2)\\
8866 (9) & & 8863.8 (5) & 10273 (10) & & 10271 (3)\\
& & 8903.5 (6) & & & 10319 (2)\\
8938 (6) & 8931 (13) & 8930 (2) & & & 10328 (3)\\
& & 8595.4 (5) & & & 10341 (3)\\
& & 9020 (2) & 10350 (7) & 10357 (14) & 10350.37 (12)\\
& & 9044.7 (3) & & & 10362.42 (7)\\
& & 9064 (1) & & & 10377 (2)\\
& & 9111 (1) & & & 10400 (15)\\
& & 9169 (1) & & & 10414 (3)\\
& & 9206 (2) & & & 10487 (3)\\
& & 9238.7 (5) & 10495 (9) & & 10493 (3)\\
& & 9261 (2) & & & 10516 (3)\\
9276 (10) & & 9281 (3) & & & 10529 (2)\\
& & 9291 (2) & & & 10567 (3)\\
& & 9304 (2) & 10575 (10) & & 10576 (2)\\
& & 9317 (2) & & & 10599.96 (7)\\ 
& & 9325.51 (6) & \\
\end{tabular}
\end{ruledtabular}
\end{table*}

\begin{table*}
\caption{The $^{26}$Mg excitation energies measured above the $\alpha$-threshold (10.615 MeV) and below the neutron threshold (11.093 MeV) in the present work along with the comparison with previous works. The numbers in parenthesis are the uncertainties in the last digits of the energy values.}
\label{tab:energy_levels_2}
\begin{ruledtabular}
\begin{tabular}{c c c c c c c c c c}
\hline 
\multicolumn{2}{c}{Present Work} & Endt98 ~\citep{Endt19981} & Massimi et al. ~\citep{PhysRevC.85.044615} & Jaeger et al.~\citep{PhysRevLett.87.202501} & Longland et al. ~\citep{PhysRevC.80.055803}\\ [0.2cm]
$^{26}$Mg($\alpha,\alpha')^{26}$Mg & $^{22}$Ne($^6$Li,d)$^{26}$Mg & Compilation & $^{25}$Mg(n,$\gamma)^{26}$Mg & $^{22}$Ne($\alpha$,n)$^{25}$Mg & $^{26}$Mg($\gamma,\gamma')^{26}$Mg \\[0.2cm] 
$E_x$(keV) & $E_x$(keV) & $E_x$(keV) & $E_x$(keV) & $E_x$(keV) & $E_x$(keV)\\ 
\hline\hline\\
& & 10646 (2) & & & 10647.3 (8)\\
& & 10650 (2) & & &\\
& & 10681.9 (3) & & &\\
& & 10693 (3) & & &\\
& & 10707 (3) & & &\\
10718 (10) & 10714 (20) & 10718.75 (9) & & &\\
& & 10726 (3) & & &\\
& & 10745.98 (12) & & &\\
& & 10767 (2) & & &\\
& & 10805.9 (4) & & & 10805.7 (7)\\
10822 (10) & & 10824 (3) & & &\\
& & 10881 (3) & & &\\
& & 10893 (3) & & &\\
& & 10915 (3) & & &\\ 
& & 10927 (3) & & &\\
10937 (11) & 10977 (15) & 10945 (3) & & & 10949.1 (8)\\
& & 10978 (3) & & &\\
& & 10998 (3) & & &\\
& & 11012 (3) & & &\\
& & 11048 (3) & & &\\
11085 (8) & & 11084 (3) & & &\\ 
\end{tabular}
\end{ruledtabular}
\end{table*} 

\begin{table*}
\caption{The $^{26}$Mg excitation energies measured above the neutron threshold (11.093 MeV) in the present work along with the comparison with previous works. The numbers in parenthesis are the uncertainties in the last digits of the energy values.}
\label{tab:energy_levels_3}
\begin{ruledtabular}
\begin{tabular}{c c c c c c c c c c}
\hline 
\multicolumn{2}{c}{Present Work} & Endt98 ~\citep{Endt19981} & Massimi et al. ~\citep{PhysRevC.85.044615} & Jaeger et al.~\citep{PhysRevLett.87.202501} & Longland et al. ~\citep{PhysRevC.80.055803}\\ [0.2cm]
$^{26}$Mg($\alpha,\alpha')^{26}$Mg & $^{22}$Ne($^6$Li,d)$^{26}$Mg & Compilation & $^{25}$Mg(n,$\gamma)^{26}$Mg & $^{22}$Ne($\alpha$,n)$^{25}$Mg & $^{26}$Mg($\gamma,\gamma')^{26}$Mg \\[0.2cm] 
$E_x$(keV) & $E_x$(keV) & $E_x$(keV) & $E_x$(keV) & $E_x$(keV) & $E_x$(keV)\\ 
\hline\hline\\
& & 11112.2 (2) & 11112.19 (9) & &\\
& & 11142 (6) & & &\\
& & 11153.2 (2) & 11153.474 (43) & & 11153.5 (10)\\
& & 11163.3 (5) & 11163.04 (7) & &\\
11167 (9) & 11169 (17) & 11169.4 (2) & 11169.42 (7) & &\\
& & 11171.1 (7) & 11171.183 (41) & &\\
& & 11183.0 (2) & 11183.20 (6) & &\\
& & 11188.8 (2) & 11189.40 (6) & &\\
& & 11191 (2) & 11191.289 (49) & &\\
& & 11194.5 (2) & & &\\
& & & 11196.68 (6) & &\\
& & 11243.3 (2) & 11243.62 (6) & &\\
& & 11274.4 (2) & 11274.441 (49) & &\\
& & 11279.5 (2) & 11280.349 (49) & &\\
& & & 11285.86 (7) & &\\
& & 11286.6 (3) & 11286.572 (46) & &\\
& & 11289.2 (3) & 11289.397 (41) & &\\
11301 (9) & & 11294.7 (5) & 11293.63 (5) & &\\
& & & 11296.39 (9) & &\\
& & 11311.0 (5) & 11310.945 (41) & &\\
& 11317 (18) & & & 11319 (2) &\\
& & 11328.3 (5) & 11326.56 (6) & &\\
& & 11329 (2) & 11328.61 (7) & &\\  
& & & 11329.527 (42) & &\\
& & & 11337.31 (5) & &\\
& & 11343.7 (5) & 11345.21 (7) & &\\  
11359 (8) & & 11362.0 (6) & 11362.31 (24) & &\\
& & 11364.9 (6) & & &\\
& & 11372.5 (6) & & &\\
& & 11392.7 (6) & 11393.10 (5) & &\\
& & 11425.4 (7) & & &\\
11445 (9) & & 11439.8 (7) & 11441.70 (6) & 11441 (2) &\\
& & 11457 (2) & & &\\
& & 11463.9 (8) & 11466.29 (8) & 11461 (2) &\\
& & 11499.4 (8) & & &\\
11509 (11) & & 11508.1 (9) & 11500.82 (5) & 11506 (2) &\\
& & 11540.8 (9) & 11527.60 (10) & 11526 (2) &\\
& & 11570 (2) & & &\\
& & 11586 (1) & 11588.88 (7) & &\\
& & 11612 (5) & 11609.22 (6) & 11630 (2) &\\
11648 (7) & & 11647 (5) & & &\\
11731 (9) & & & & 11749 (10) &\\
& & 11795 (10) & & 11787 (4) &\\
11824 (9) & & 11828 (3) & & 11828 (2) &\\
& & 11890 (2) & & &\\
11900(9) & & 11910 (2) & & &\\
& & 11945 (10) & & &\\
& & 11950 (2) & & &\\
12064 (8) & & 12049 (2) & & &\\
\end{tabular}
\end{ruledtabular}
\end{table*}  

\subsection{Angular Distribution Analysis}

\begin{figure*}
   \includegraphics[width=2\columnwidth]{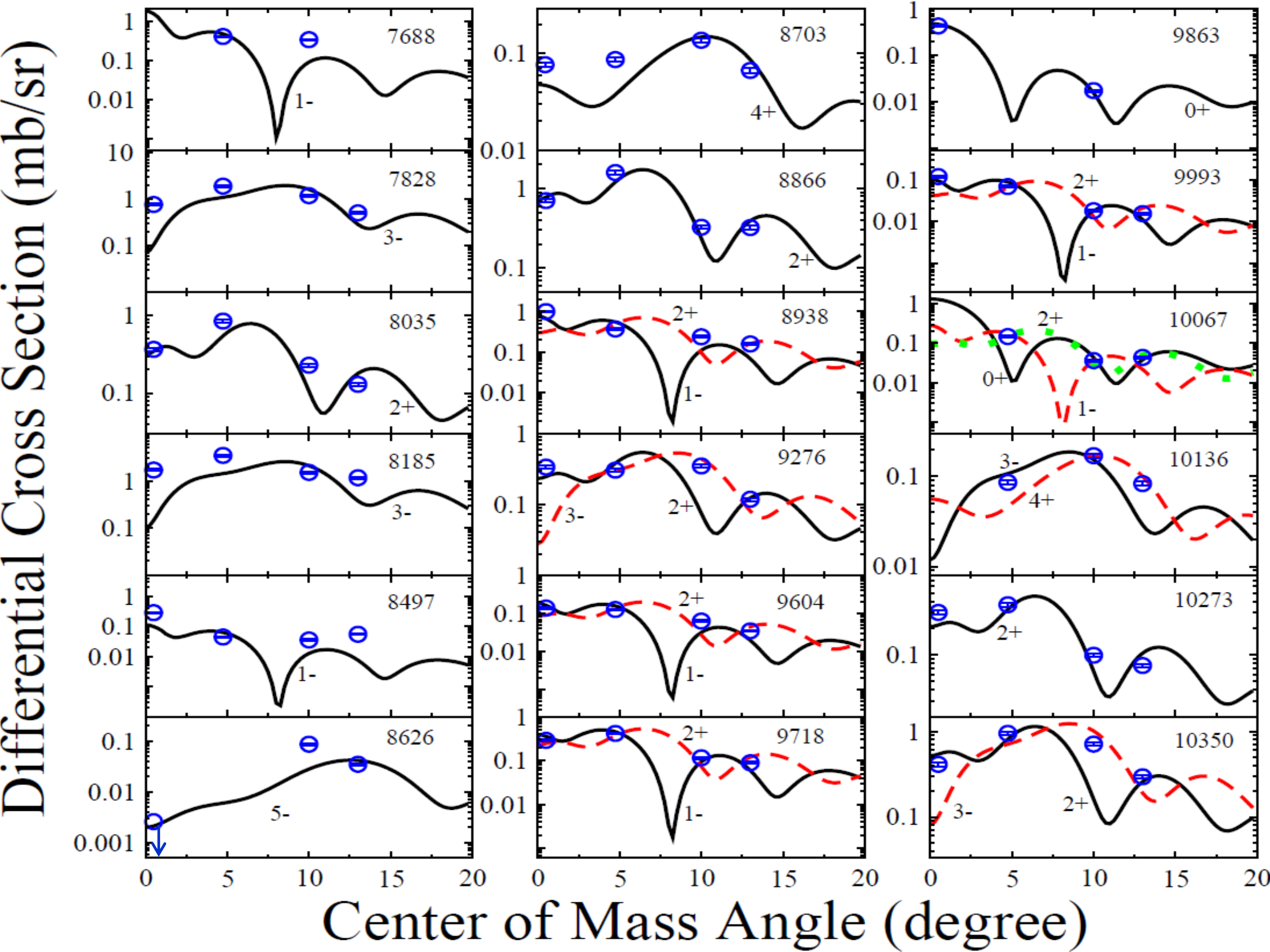}
    \caption{(Color online) Angular distributions obtained using PTOLEMY for states excited in the $^{26}$Mg($\alpha,\alpha')^{26}$Mg reaction at $E_{\alpha}$ = 206 MeV. The blue circles with error bars represent the experimental data points. The empty blue circle with the blue arrow in the downward direction represents the upper limit for the cross-section at that angle. The resulting spin-parities are based on present distributions and values quoted in literature.}
    \label{fig:26Mg_Plots_1}
\end{figure*}

\begin{figure*}
   \includegraphics[width=2\columnwidth]{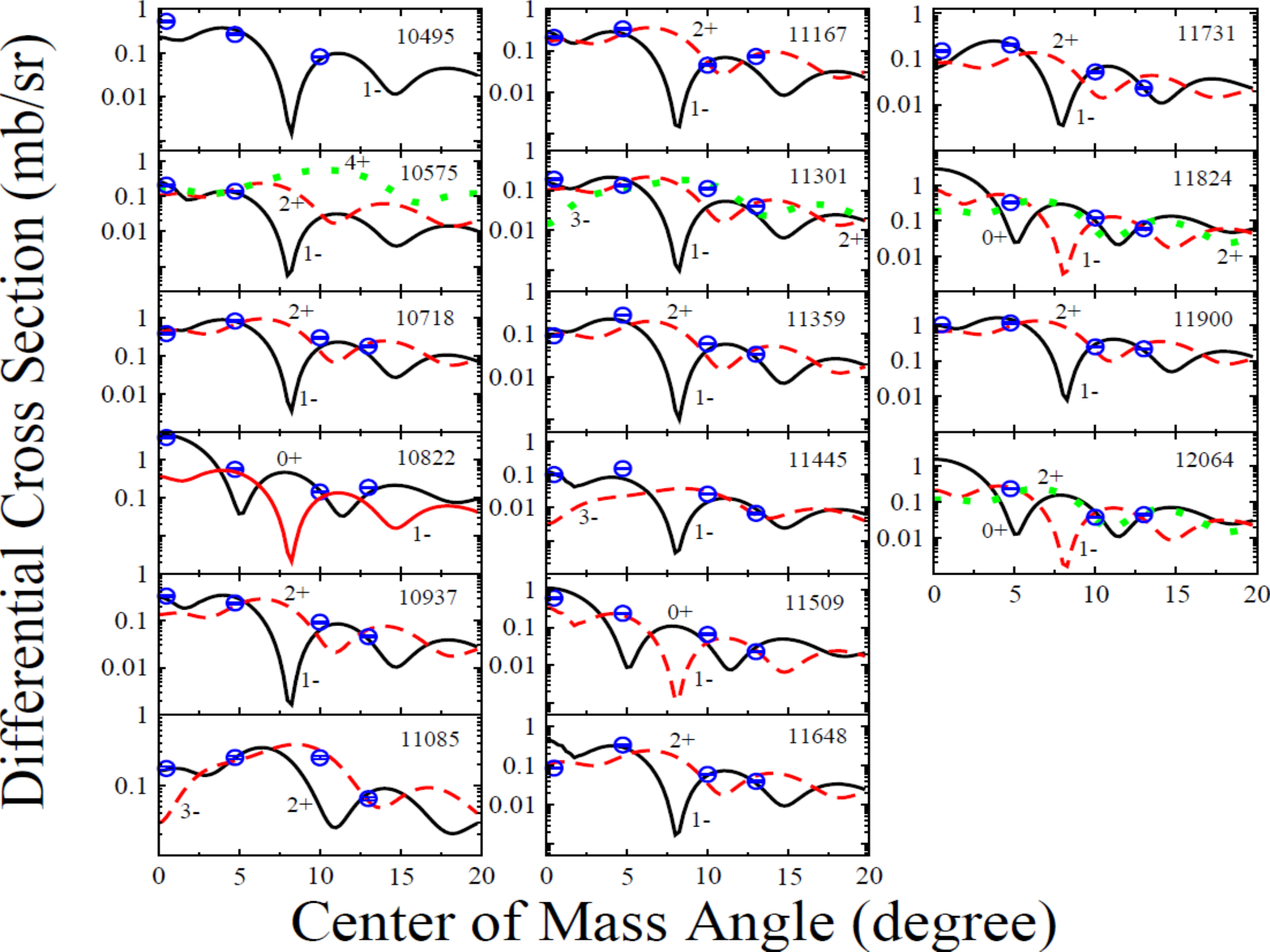}
    \caption{(Color online) Continuation of angular distributions obtained using PTOLEMY for states excited in the $^{26}$Mg($\alpha,\alpha')^{26}$Mg reaction at $E_{\alpha}$ = 206 MeV. The blue circles with error bars represent the experimental data points. The resulting spin-parities are based on present distributions and values quoted in literature. }
    \label{fig:26Mg_Plots_2}
\end{figure*}

\begin{figure*}
  \includegraphics[width=2\columnwidth]{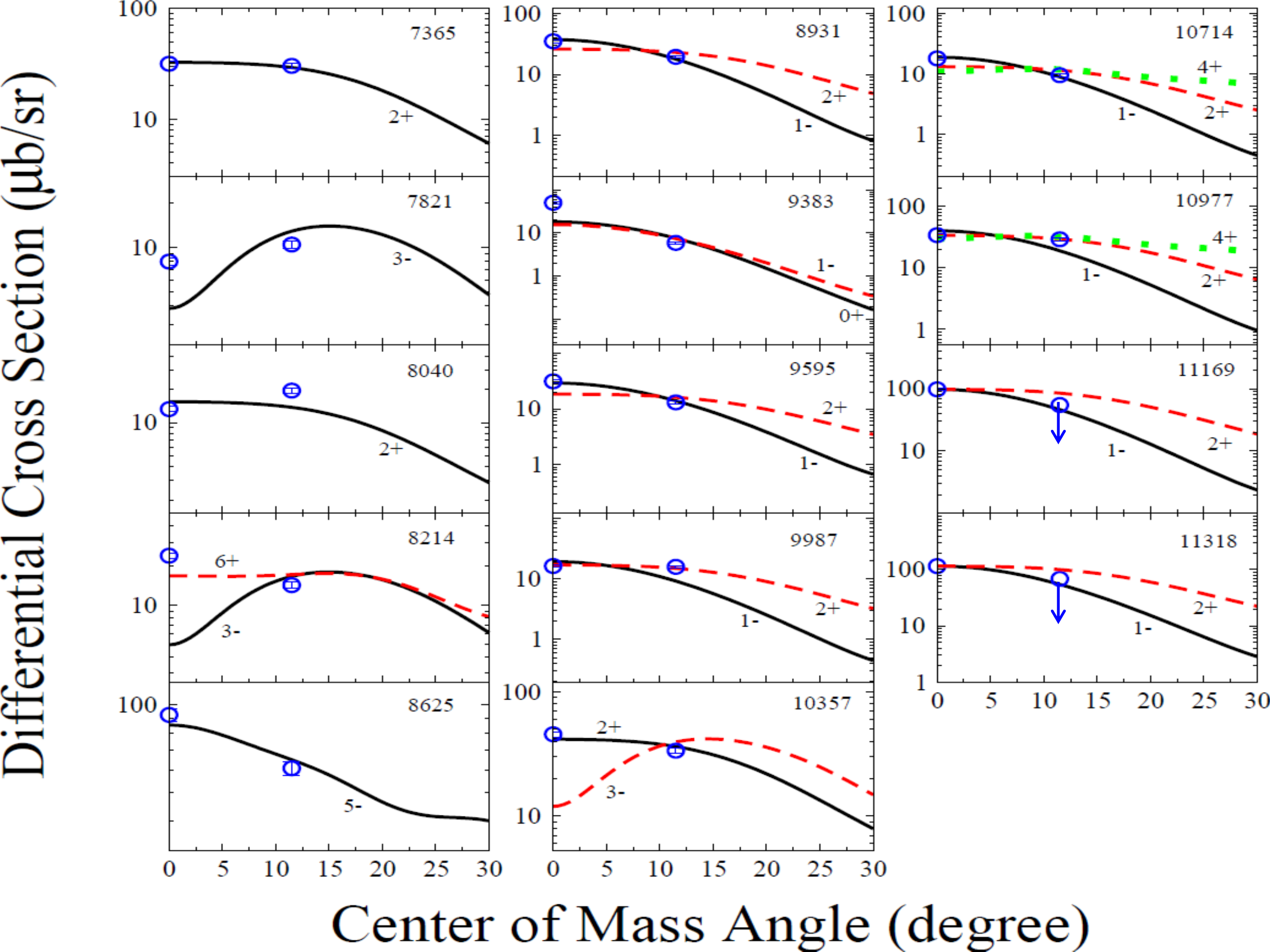}
    \caption{(Color Online) Angular distributions obtained using FRESCO for states excited in the $^{22}$Ne($^6$Li,d)$^{26}$Mg reaction at $E_{Li}$ = 82.7 MeV. The blue circles with error bars represent the experimental data points. The empty blue circles with the blue arrows in the downward direction represent the upper limit for the cross-section at that angle. The resulting spin-parities are based on present distributions and values quoted in literature.}
    \label{fig:22Ne_Plots_1}
 \end{figure*}

The angular distributions in the present work were studied using the general purpose inelastic coupled channel code called PTOLEMY ~\citep{PTOLEMY} for ($\alpha,\alpha'$) and the state-of-the-art code for transfer reactions called FRESCO ~\citep{FRESCO} for ($^6$Li,$d$), under the assumption that the observed peaks are the result of a single level in $^{26}$Mg. The starting set of optical potential parameters were adopted from references ~\citep{VanDerBorg1981243}, ~\citep{Cook1982153} and ~\citep{PhysRevC.73.054605} and were then modified to best fit the present $^{26}$Mg($\alpha,\alpha'$) and $^{22}$Ne($^6$Li,d) data. The final set of optical parameters are given in Tables ~\ref{tab:parameters_for_alpha} and ~\ref{tab:parameters_for_6Li,d} . For the alpha transfer study, a Woods-Saxon potential of radius $r$~=~ 1.31$A_T^{1/3}$ and a diffuseness $a$ = 0.65 was used. The number of radial nodes N and the orbital momentum L were fixed by the Talmi-Moshinsky relation, $\sum\limits_{i=1}^4 (2n_i + l_i)$, where $n_i,l_i$ refer to the harmonic oscillator quantum numbers of each transferred nucleon ~\citep{Giesen199395}. For all positive parity states (L = even), the (sd)$^4$ configuration was assumed resulting in 2N+L = 8 and for all negative parity states (L = odd), the (sd)$^3$(fp) configuration was assumed giving 2N+L = 9. Figures ~\ref{fig:26Mg_Plots_1}, ~\ref{fig:26Mg_Plots_2}, and ~\ref{fig:22Ne_Plots_1} show the resulting angular distributions for the ($\alpha,\alpha'$) and the ($^6$Li,$d$) measurements, respectively. 

\begin{table*}
\caption{Optical parameters used in PTOLEMY to study the angular distributions of $^{26}$Mg($\alpha,\alpha')^{26}$Mg cross-sections.}
\label{tab:parameters_for_alpha}
\begin{ruledtabular}
\begin{tabular}{c c c c c c c c c}
Nucleus & $E_{\alpha}$ & V & $r_{0R}$ & $a_R$ & $V_I$ & $r_{0I}$ & $a_I$ & $r_{0C}$ \\
& (MeV) & (MeV) & (fm) & (fm) & (MeV) & (fm) & (fm) & (fm) \\ 
\hline
$^{26}$Mg & 206 & 100.0 & 1.20 & 0.61 & 25.67 & 1.50 & 0.55 & 1.30 \\
\end{tabular}
\end{ruledtabular}
\end{table*} 

\begin{table*}
\caption{Optical parameters used in FRESCO for DWBA analysis of $^{22}$Ne($^6$Li,d)$^{26}$Mg.}
\label{tab:parameters_for_6Li,d}
\begin{ruledtabular}
\begin{tabular}{c c c c c c c c c}
Reaction Channel & V & $r_{0R}$ & $a_R$ & $W_s$ & $4W_D$ & $r_{0I}$ & $a_I$ & $r_{0C}$ \\
& (MeV) & (fm) & (fm) & (MeV) & (MeV) & (fm) & (fm) & (fm) \\ 
\hline
$^{22}$Ne + $^6$Li & 109.50 & 1.33 & 0.81 & 51.30 & & 1.53 & 0.88 & 1.23 \\
$^{26}$Mg + d & 72.90 & 1.16 & 0.76 & & 8.10 & 1.34 & 0.56 & 1.30 \\
final state & $^a$ & 1.31 & 0.65 \\
\end{tabular}
\end{ruledtabular}
\footnotetext[1]{Adjusted to give the correct binding energy.}
\end{table*} 

\subsection{Discussion of peaks above the $\alpha$-threshold} \label{subsec:discussion}

Above the $\alpha$-threshold (10.615 MeV), the following peaks have been observed in the region of interest ($E_x$ = 10.61 - 11.32 MeV) : \textbf{$E_x$ = 10717 (9), 10822 (10), 10951 (21), 11085 (8), 11167 (8) and 11317 (18) keV}. These energies are weighted averages of the energies measured in the present ($\alpha,\alpha'$) and ($^6$Li,d) experiments. Peaks corresponding to all of these states have been seen in the ($\alpha,\alpha'$) experiment at all four angles, 0.45$^{\circ}$, 4.1$^{\circ}$, 8.6$^{\circ}$ and 11.1$^{\circ}$, except for the $E_x$ = 11317 (18) keV state. This state could not be clearly identified because it was partly obscured by the $E_x$ = 11301 (9) keV state and partly by the $E_x$ = 11359 (8) keV state in $^{26}$Mg. In the ($^6$Li,$d$) experiment, the above mentioned six peaks were observed at 0$^{\circ}$ and 10$^{\circ}$ except for the $E_x$ = 11167 (8) keV state and the $E_x$ = 11317 (18) keV state which were observed only at 0$^{\circ}$. At 10$^{\circ}$, the $E_x$ = 11167 (8) keV peak was partly covered by the $E_x$ = 9532.48 (10) keV state ~\citep{Firestone20072319} in $^{24}$Mg and  the $E_x$ = 11317 (18) keV peak was partly covered by the $E_x$ = 9532.48 (10) keV state ~\citep{Firestone20072319} in $^{24}$Mg and partly by the $E_x$ = 4247.7 (11) keV state ~\citep{Tilley1998249} in $^{20}$Ne. As can be seen in Tables ~\ref{tab:energy_levels_2} and ~\ref{tab:energy_levels_3}, the energies in the present work are in good agreement with those from previous transfer measurements with similar energy resolution. The spin parity possibilities for these peaks as well as the final adopted values are tabulated in Table ~\ref{tab:spin-parity}. Unlike Giesen et al. ~\citep{Giesen199395}, the excited states observed in the present work correspond to lower angular momentum transfer. The $E_x$ = 11167 (8) keV state and the $E_x$ = 11317 (18) keV state require a more detailed discussion.\\

\begin{table*}
\caption{Spin-parity assignments for states populated above the $\alpha$-threshold in the present ($\alpha,\alpha'$) and ($^6$Li,$d$) experiments. The states mentioned above the line are open only in the $\alpha$-channel and the ones below are open in both the $\alpha$- channel and the $n$-channel.}
\label{tab:spin-parity}
\begin{ruledtabular}
\begin{tabular}{c c c c c c c c c c c}
$E_x$ & $E_R^{c.m.}$  & \multicolumn{4}{c}{$J^{\pi}$}\\ [0.2cm]
\cline{3-7}
& & \multicolumn{2}{c}{Present Work} & Longland et al. ~\citep{PhysRevC.80.055803} & Giesen et al. ~\citep{Giesen199395} & Adopted value(s) \\
(keV) & (keV) & $^{26}$Mg($\alpha,\alpha')^{26}$Mg & $^{22}$Ne($^6$Li,d)$^{26}$Mg & $^{26}$Mg($\gamma,\gamma')^{26}$Mg & $^{22}$Ne($^6$Li,d)$^{26}$Mg & \\
\hline\\
10717 (9) & 102 & 1$^-$,2$^+$ & 1$^-$,2$^+$,4$^+$ & & 4$^+$,7$^-$,8$^+$ & 1$^-$,2$^+$\\
10822 (10) & 207 & 0$^+$,1$^-$ & & 1$^-$ & & 1$^-$ \\
10951 (21) & 336 & 1$^-$,2$^+$ & 1$^-$,2$^+$,4$^+$ & 1$^-$ & (2$^+$,4$^+$),3$^-$ & 1$^-$ \\  
11085 (8) & 471 & 2$^+$,3$^-$ & & & & 2$^+$,3$^-$ \\
\hline 
11167 (8) & 553 & 1$^-$,2$^+$ & 1$^{-(a)}$ & & & 1$^-$ \\ 
11317 (18) & 702 & & 1$^{-(a)}$ & & (1$^-$),2$^+$ & 1$^-$  \\
\end{tabular}
\end{ruledtabular}
\footnotetext[1]{These values are based on the upper limit for the ($^6$Li,$d$) cross-section determined at 10$^{\circ}$.}
\end{table*}

\textbf{$E_x$ = 11167 (8) keV ($E_R$ = 553 keV; $J^{\pi}$ = 1$^-$):} The cross-sections from the ($\alpha,\alpha'$) experiment follow the 1$^-$ as well as the 2$^+$ angular distributions. But, based on the upper limit derived for this state at 10$^{\circ}$ in the ($^6$Li,$d$) measurement, the data  favor a 1$^-$ angular distribution. Hence, a spin-parity of 1$^-$ was assigned to this state. As can be seen in Table ~\ref{tab:energy_levels_3}, Massimi et al. ~\citep{PhysRevC.85.044615} has reported five additional neutron resonances within $\pm$20 keV of the $E_x$ = 11167 keV level observed in the present measurements. Among these five resonances, only the 1$^-$ level at $E_x$ = 11183.20 keV ($E_n$ = 92.60 (2) keV) matches the angular distribution determined in the present work. Hence, the neutron and gamma widths (shown in Table ~\ref{tab:widths}) were adopted corresponding to this 1$^-$ state at $E_n$ = 93.60 (2) keV for the reaction rate calculations. 

\textbf{$E_x$ = 11317 (18) keV ($E_R$ = 702 keV; $J^{\pi}$ = 1$^-$):} Koehler ~\citep{PhysRevC.66.055805} has made an argument that this state cannot correspond to both the $^{22}$Ne($\alpha$,n)$^{25}$Mg resonance observed at $E_R^{lab}$ = 832 (2) keV ($E_x$ = 11319 (2) keV) ~\citep{PhysRevLett.87.202501} and the $^{22}$Ne($\alpha,\gamma)^{26}$Mg resonance observed at $E_R^{lab}$ = 828 (5) keV ($E_x$ = 11315 (5) keV) ~\citep{Wol89}. The basis of his argument is the assumption that for $E_R^{lab}$ = 832 keV, the total width $\Gamma$ = 0.25 (0.17) keV, as reported  by reference ~\citep{PhysRevLett.87.202501}. Since the energy resolution for the Stuttgart DYNAMITRON accelerator, that was used for the $^{22}$Ne($\alpha$,n)$^{25}$Mg measurement by Jaeger et al. ~\citep{PhysRevLett.87.202501} , is 1.4 keV ~\citep{Hammer1979189}, the reported $\Gamma$ value of 0.25 keV can only be treated as an upper limit and not the actual value. Hence, the basis of Koehler's argument is incorrect. The (n,$\gamma$) measurements ~\citep{PhysRevC.85.044615}, ~\citep{PhysRevC.14.1328} have seen four resonances at $E_n$ = 226.19, 242.45, 244.58 and 245.57 keV. None of these correspond to the well known $E_R^{lab}$ = 832 $\pm$ 2 keV ($E_n$ = 235 $\pm$ 2 keV) resonance, within error bars, which has been observed in references ~\citep{PhysRevLett.87.202501}, ~\citep{Wol89} and ~\citep{PhysRevC.43.2849}. Also, the 702 keV (E$_R^{lab}$ = 830 keV) resonance, observed in the present work, has a pronounced $\alpha$-cluster structure, as reflected by its large $\alpha$-spectroscopic factor (Table ~\ref{tab:widths}) with a $\Gamma_{\gamma}$ / $\Gamma_n$ ratio = 0.3 (determined using $\omega\gamma_{(\alpha,\gamma)}$ = 0.036 (4) meV ~\citep{Wol89} and $\omega\gamma_{(\alpha,n)}$ = 0.118 (11) meV ~\citep{PhysRevLett.87.202501}. This implies that the neutron width associated with this resonance should be small, and, therefore the probability of observing it in an (n,$\gamma$) measurement is low. Besides, it is highly unlikely that two pronounced $\alpha$ cluster states should be observed right next to each other, as per Koehler's claim. 

\subsection{Reaction Rates}  \label{subsec:rate}

The alpha capture rates on $^{22}$Ne have been determined using the narrow resonance reaction rate formalism defined as ~\citep{Angulo19993}:

\begin{multline}
N_A\langle\sigma\nu\rangle = 1.54\times 10^5(\mu T_9)^{-3/2} \\  \sum\limits_i (\omega\gamma)_i \exp\left(\frac{-11.605E_{R,i}}{T_9}\right) \\  cm^3 sec^{-1} mol^{-1}
\end{multline}

\noindent where, $\mu$ is the reduced mass, $T_9$ is the temperature in GK, $(\omega\gamma)_i$ is the resonance strength of the i$^{th}$ resonance in eV and $E_{R,i}$ is the resonance energy in the center of mass frame of the i$^{th}$ resonance in MeV.

The resonance energies were determined using $E_{R,i} = E_{x,i} - Q$(10.6150 (2) MeV) and the resonance strengths were calculated using the following ~\citep{Angulo19993} :

\begin{equation}
\omega\gamma_{(\alpha,\gamma)} = \frac{2J+1}{(2J_1+1)(2J_2+1)}\frac{\Gamma_{\alpha}\Gamma_{\gamma}}{\Gamma}
\end{equation}
and 
\begin{equation}
\omega\gamma_{(\alpha,n)} = \frac{2J+1}{(2J_1+1)(2J_2+1)}\frac{\Gamma_{\alpha}\Gamma_n}{\Gamma}
\end{equation}

\noindent where, $J$ represents the spin of the resonance and $J_1$ and $J_2$ represent the spin of $^{22}$Ne and $\alpha$ nuclei, respectively, both being equal to 0. Because of the penetrability, for low energy resonances $\Gamma_{\alpha} \ll \Gamma_{\gamma}$ and $\Gamma_n$ (~\citep{Giesen199395},~\citep{Wol89}). For n-bound states ($\Gamma$ = $\Gamma_{\alpha}$ + $\Gamma_{\gamma}$), Eq. (3) can be written as :

\begin{equation}
\omega\gamma_{(\alpha,\gamma)} = (2J+1)\Gamma_{\alpha}
\end{equation}

\noindent For n-unbound states ($\Gamma$ = $\Gamma_{\alpha}$ + $\Gamma_{\gamma}$ + $\Gamma_n$), Eq. (3) and Eq. (4) can be written as :

\begin{equation}
\omega\gamma_{(\alpha,\gamma)} = (2J+1)\frac{\Gamma_{\alpha}\Gamma_{\gamma}}{\Gamma_{\gamma}+\Gamma_n}
\end{equation}
and 
\begin{equation}
\omega\gamma_{(\alpha,n)} = (2J+1)\frac{\Gamma_{\alpha}\Gamma_n}{\Gamma_{\gamma}+\Gamma_n}.
\end{equation}

The ($^6$Li,$d$) angular distributions obtained using FRESCO were used to compute the relative alpha spectroscopic factors ($S_{\alpha (rel)}$) using the following equation:    
\begin{equation}
\frac{d\sigma_{exp}}{d\Omega} = S_{\alpha (rel)} N \sigma_{DWBA}
\end{equation}
where, N is the normalization constant. For ($^6$Li,$d$) and ($d$,$^6$Li) reactions, N has been found ~\citep{Milder197772} to be equal to 2.67 from a comparison of $\alpha$-transfer and $\alpha$-decay data ~\citep{Giesen199395}. The $S_{\alpha (rel)}$ values determined using the above equation were scaled by a factor of 2 to reproduce the $\alpha$-width ($\Gamma_{\alpha}$) corresponding to the lowest directly observed resonance at $E_R^{lab}$ = 832 keV. The resulting alpha spectroscopic factors were then used to determine the alpha widths for the present measurement using :
\begin{equation}
\Gamma_{\alpha} = S_{\alpha}\Gamma_{sp}.
\end{equation}

\noindent where $\Gamma_{sp}$ represent the single particle widths calculated using the optical potential parameters listed in Table ~\ref{tab:parameters_for_6Li,d}. 

The resulting resonance parameters for the levels observed in the astrophysical region of interest have been listed in Table ~\ref{tab:widths}. Using these parameters the $\alpha$-capture rates were calculated for the present work, as shown in Tables ~\ref{tab:rate_ag} and ~\ref{tab:rate_an}. As discussed in section ~\ref{subsec:discussion}, the neutron and  gamma widths for the neutron unbound level at $E_x$ = 11167 keV have been adopted from Ref. ~\citep{PhysRevC.85.044615}. However, using the values of $\Gamma_{\gamma}$ = 2 (2) eV and $\Gamma_n$ = 0.6 (0.2) eV from Ref. ~\citep{PhysRevC.85.044615}, the value for $\omega\gamma_{(\alpha,n)}$ is equal to 125 neV. This value is higher than the upper limit of 60 neV obtained by Jaeger et al ~\citep{PhysRevLett.87.202501}. Hence, for this state, the ($\alpha$,n) rate contribution was calculated using $\omega\gamma_{(\alpha,n)}$ = 60 neV, the upper limit was calculated using $\omega\gamma_{(\alpha,n)}$ = 125 neV and the lower limit was determined by maximizing the $\Gamma_{\gamma}$ / $\Gamma_n$ ratio (i.e. adopting $\Gamma_{\gamma}$ = 4 eV and $\Gamma_n$ = 0.4 eV ~\citep{PhysRevC.85.044615}), within error bars. Because $\Gamma_{\gamma}$ $\gg$ $\Gamma_n$, the ($\alpha,\gamma$) rate contribution was calculated using Eq. 4. The $E_x$ = 10717 keV level had a negligible contribution to the alpha capture rates in the temperature window of interest (0.01 $\textless$ $T_9$ $\textless$ 10), and hence has not been included in the total reaction rate calculations.  

\begin{table*}
\caption{Resonance parameters for the $^{26}$Mg resonances observed in the present measurements in the astrophysical region of interest. The $S_{\alpha}$ values listed here were obtained by scaling the relative $\alpha$-spectroscopic factors ($S_{\alpha (rel)}$) by a factor of 2, as explained in section ~\ref{subsec:rate} }
\label{tab:widths}
\begin{ruledtabular}
\begin{tabular}{c c c c c c c c c c}
$E_x$ & $E_R^{c.m.}$  & $J^{\pi}$ & $S_{\alpha}$ & $\Gamma_{sp}$ & $\Gamma_{\alpha}$ & $\Gamma_{\gamma}^a$ & $\Gamma_n^a$ & $\omega\gamma_{(\alpha,\gamma)}$ & $\omega\gamma_{(\alpha,n)}$\\ [0.2cm]
(keV) & (keV) & & & (eV) & (eV) & (eV) & (eV) & (eV) & (eV) \\
\hline\hline\\
10717 (9) & 102 & 1$^-$ & 0.07 & 3.78$\times$10$^{-35}$ & 2.8 (2)$\times$10$^{-36}$ & & & 8.5 (5)$\times$10$^{-36}$ &\\
& & 2$^+$ & 0.14 & 6.00$\times$10$^{-36}$ & 9 (2)$\times$10$^{-37}$ & & & 4 (1)$\times$10$^{-36}$ &\\[0.2cm]
10822$^b$ (10) & 207 & 1$^-$ & $\leq$0.07 & $\leq$2.99$\times$10$^{-20}$ & $\leq$1.97$\times$10$^{-21}$ & & & $\leq$5.92$\times$10$^{-21}$ &\\[0.2cm]
10951 (21) & 336 & 1$^-$ & 0.16 & 5.68$\times$10$^{-13}$ & 9 (3)$\times$10$^{-14}$ & & & 2.8 (8)$\times$10$^{-13}$ &\\[0.2cm]
11085$^b$ (8) & 471 & 2$^+$ & $\leq$0.07 & $\leq$7.01$\times$10$^{-11}$ & $\leq$4.71$\times$10$^{-12}$ & & & $\leq$2.36$\times$10$^{-11}$ &\\
& & 3$^-$ & $\leq$0.07 & $\leq$9.77$\times$10$^{-12}$ & $\leq$7.06$\times$10$^{-13}$ & & & $\leq$4.95$\times$10$^{-12}$ &\\[0.2cm]
11167 (8) & 553 & 1$^-$ & 0.40 & 5.00$\times$10$^{-07}$ & 2.0 (1)$\times$10$^{-07}$ & 2 (2) & 0.6 (0.2) & 5.5 (4)$\times$10$^{-07}$ & 6$\times$10$^{-08}$ $^c$\\[0.2cm]
11317 (18) & 702 & 1$^-$ & 0.48 & 1.05$\times$10$^{-04}$ & 5.0 (3)$\times$10$^{-05}$ & & & 3.7 (4)$\times$10$^{-05}$ $^d$ & 1.2 (1)$\times$10$^{04}$ $^d$ \\
\end{tabular}
\end{ruledtabular}
\footnotetext[1]{These values have been adopted from ~\citep{PhysRevC.85.044615}}
\footnotetext[2]{These peaks have not been seen in $^{22}Ne(^6Li,d)^{26}Mg$ spectra. However they were seen in $^{26}Mg(\alpha,\alpha')^{26}Mg$ spectra. Hence, using kinematics, their positions were determined in the ($^6$Li,$d$) spectra and upper limits for their corresponding resonance parameters were determined.}
\footnotetext[3]{This value is the upper limit predicted by Jaeger et al. ~\citep{PhysRevLett.87.202501}}
\footnotetext[4]{These values have been adopted from ~\citep{Wol89} and ~\citep{PhysRevLett.87.202501}}
\end{table*}

\begingroup
\squeezetable
\begin{table*}
\caption{Monte Carlo reaction rates for the $^{22}$Ne($\alpha,\gamma)^{26}$Mg reaction calculated using the Nucleosynthesis Simulator called Starlib ~\citep{starlib}. The median rate represents the recommended ($\alpha,\gamma$) rate determined using the contributions from the 336 keV ($E_x$ = 10951 keV), the 553 keV ($E_x$ = 11167 keV) and the 702 keV ($E_x$ = 11317 keV) resonances observed in the present work along with the other known resonances reported in the literature from the direct measurement of the $^{22}$Ne($\alpha,\gamma)^{26}$Mg reaction. The contribution from the 102 keV ($E_x$ = 10717 keV) resonance was negligible and hence has not been included in the rate calculation. The upper limit contributions from the 207 keV (E$_x$ = 10822 keV) and the 2$^+$ 471 keV ($E_x$ = 11085 keV) resonances, seen only in the present ($\alpha,\alpha'$) experiment, have been added to the high rate. The contribution from the 471 keV ($E_x$ = 11085 keV) resonance corresponding to the 3$^-$ angular distribution was added to the low rate. The rate values in parenthesis represent the temperatures (T$_9$ $\textgreater$ T$_{9 match}$ = 1.5) for which Non-Smoker Hauser Feshbach rates from the JINA Reaclib Database ~\citep{Jin10}, normalized to the experimental results, have been adopted. The Longland et al. ~\citep{PhysRevC.85.065809} and NACRE ~\citep{Angulo19993} rates have also been provided to facilitate the comparison. }
\label{tab:rate_ag}
\begin{ruledtabular}
\begin{tabular}{c c c c c c c c c c}
& \multicolumn{3}{c}{Present Work} & \multicolumn{3}{c}{Longland et al. ~\citep{PhysRevC.85.065809}} & \multicolumn{3}{c}{NACRE ~\citep{Angulo19993}}\\
$T_9$ & Low rate  & Median rate & High rate & Low rate  & Median rate & High rate & Low rate  & Median rate & High rate\\ 
\hline\\

0.01 & 4.91$\times$10$^{-81}$ & 6.20$\times$10$^{-81}$ & 8.01$\times$10$^{-81}$ & 1.05$\times$10$^{-77}$ & 2.14$\times$10$^{-77}$ & 4.52$\times$10$^{-77}$ & 0.00 & 0.00 & 0.00\\
0.011 & 4.26$\times$10$^{-78}$ & 5.37$\times$10$^{-78}$ & 6.94$\times$10$^{-78}$ & 3.99$\times$10$^{-74}$ & 7.28$\times$10$^{-74}$ & 1.34$\times$10$^{-73}$ & 0.00 & 0.00 & 0.00\\
0.012 & 1.70$\times$10$^{-75}$ & 2.15$\times$10$^{-75}$ & 2.77$\times$10$^{-75}$ & 3.69$\times$10$^{-71}$ & 6.34$\times$10$^{-71}$ & 1.07$\times$10$^{-70}$ & 0.00 & 0.00 & 0.00\\
0.013 & 3.61$\times$10$^{-73}$ & 4.55$\times$10$^{-73}$ & 5.88$\times$10$^{-73}$ & 1.15$\times$10$^{-68}$ & 1.90$\times$10$^{-68}$ & 3.09$\times$10$^{-68}$ & 0.00 & 0.00 & 0.00\\
0.014 & 4.54$\times$10$^{-71}$ & 5.72$\times$10$^{-71}$ & 7.40$\times$10$^{-71}$ & 1.55$\times$10$^{-66}$ & 2.52$\times$10$^{-66}$ & 4.04$\times$10$^{-66}$ & 0.00 & 0.00 & 0.00\\
0.015 & 3.67$\times$10$^{-69}$ & 4.63$\times$10$^{-69}$ & 5.98$\times$10$^{-69}$ & 1.06$\times$10$^{-64}$ & 1.73$\times$10$^{-64}$ & 2.79$\times$10$^{-64}$ & 0.00 & 0.00 & 0.00\\
0.016 & 2.04$\times$10$^{-67}$ & 2.57$\times$10$^{-67}$ & 3.33$\times$10$^{-67}$ & 4.11$\times$10$^{-63}$ & 6.96$\times$10$^{-63}$ & 1.14$\times$10$^{-62}$ & 0.00 & 0.00 & 0.00\\
0.018 & 2.51$\times$10$^{-64}$ & 3.16$\times$10$^{-64}$ & 4.08$\times$10$^{-64}$ & 1.80$\times$10$^{-60}$ & 3.26$\times$10$^{-60}$ & 5.63$\times$10$^{-60}$ & 0.00 & 0.00 & 0.00\\
0.02 & 1.15$\times$10$^{-61}$ & 1.45$\times$10$^{-61}$ & 1.87$\times$10$^{-61}$ & 2.24$\times$10$^{-58}$ & 4.34$\times$10$^{-58}$ & 8.04$\times$10$^{-58}$ & 0.00 & 0.00 & 0.00\\
0.025 & 2.50$\times$10$^{-56}$ & 3.14$\times$10$^{-56}$ & 1.16$\times$10$^{-55}$ & 1.54$\times$10$^{-54}$ & 3.14$\times$10$^{-54}$ & 6.30$\times$10$^{-54}$ & 0.00 & 0.00 & 0.00\\
0.03 & 2.95$\times$10$^{-52}$ & 3.72$\times$10$^{-52}$ & 5.08$\times$10$^{-49}$ & 2.82$\times$10$^{-50}$ & 3.35$\times$10$^{-49}$ & 1.30$\times$10$^{-48}$ & 0.00 & 0.00 & 0.00\\
0.04 & 2.66$\times$10$^{-46}$ & 3.40$\times$10$^{-46}$ & 1.60$\times$10$^{-40}$ & 1.81$\times$10$^{-42}$ & 2.31$\times$10$^{-41}$ & 8.91$\times$10$^{-41}$ & 0.00 & 0.00 & 0.00\\
0.05 & 1.04$\times$10$^{-41}$ & 8.75$\times$10$^{-41}$ & 1.86$\times$10$^{-35}$ & 8.51$\times$10$^{-38}$ & 1.08$\times$10$^{-36}$ & 4.17$\times$10$^{-36}$ & 0.00 & 0.00 & 0.00\\
0.06 & 3.20$\times$10$^{-36}$ & 2.70$\times$10$^{-35}$ & 4.25$\times$10$^{-32}$ & 1.05$\times$10$^{-34}$ & 1.34$\times$10$^{-33}$ & 5.14$\times$10$^{-33}$ & 0.00 & 0.00 & 0.00\\
0.07 & 4.64$\times$10$^{-32}$ & 2.29$\times$10$^{-31}$ & 1.12$\times$10$^{-29}$ & 1.95$\times$10$^{-32}$ & 2.12$\times$10$^{-31}$ & 8.04$\times$10$^{-31}$ & 0.00 & 0.00 & 0.00\\
0.08 & 5.92$\times$10$^{-29}$ & 1.98$\times$10$^{-28}$ & 1.16$\times$10$^{-27}$ & 2.76$\times$10$^{-30}$ & 1.14$\times$10$^{-29}$ & 3.67$\times$10$^{-29}$ & 0.00 & 0.00 & 0.00\\
0.09 & 1.52$\times$10$^{-26}$ & 3.68$\times$10$^{-26}$ & 8.94$\times$10$^{-26}$ & 1.76$\times$10$^{-28}$ & 6.30$\times$10$^{-28}$ & 1.35$\times$10$^{-27}$ & 0.00 & 0.00 & 0.00\\
0.1 & 1.26$\times$10$^{-24}$ & 2.35$\times$10$^{-24}$ & 3.97$\times$10$^{-24}$ & 4.79$\times$10$^{-27}$ & 2.28$\times$10$^{-26}$ & 6.55$\times$10$^{-26}$ & 0.00 & 0.00 & 0.00\\
0.11 & 4.49$\times$10$^{-23}$ & 6.83$\times$10$^{-23}$ & 9.78$\times$10$^{-23}$ & 8.17$\times$10$^{-26}$ & 5.95$\times$10$^{-25}$ & 1.86$\times$10$^{-24}$ & 0.00 & 0.00 & 0.00\\
0.12 & 8.08$\times$10$^{-22}$ & 1.11$\times$10$^{-21}$ & 1.52$\times$10$^{-21}$ & 1.11$\times$10$^{-24}$ & 9.63$\times$10$^{-24}$ & 3.07$\times$10$^{-23}$  & 3.70$\times$10$^{-23}$ & 5.24$\times$10$^{-22}$ & 5.81$\times$10$^{-21}$\\
0.13 & 8.85$\times$10$^{-21}$ & 1.22$\times$10$^{-20}$ & 1.63$\times$10$^{-20}$ & 1.23$\times$10$^{-23}$ & 1.03$\times$10$^{-22}$ & 3.28$\times$10$^{-22}$ & 4.10$\times$10$^{-22}$ & 5.77$\times$10$^{-21}$ & 6.32$\times$10$^{-20}$\\
0.14 & 6.67$\times$10$^{-20}$ & 9.65$\times$10$^{-20}$ & 1.33$\times$10$^{-19}$ & 1.38$\times$10$^{-22}$ & 8.23$\times$10$^{-22}$ & 2.50$\times$10$^{-21}$ & 3.20$\times$10$^{-21}$ & 4.52$\times$10$^{-20}$ & 4.91$\times$10$^{-19}$\\
0.15 & 4.11$\times$10$^{-19}$ & 6.17$\times$10$^{-19}$ & 8.68$\times$10$^{-19}$ & 1.53$\times$10$^{-21}$ & 5.57$\times$10$^{-21}$ & 1.51$\times$10$^{-20}$ & 1.90$\times$10$^{-20}$ & 2.73$\times$10$^{-19}$ & 2.95$\times$10$^{-18}$\\
0.16 & 2.36$\times$10$^{-18}$ & 3.49$\times$10$^{-18}$ & 4.91$\times$10$^{-18}$ & 1.41$\times$10$^{-20}$ & 3.79$\times$10$^{-20}$ & 8.10$\times$10$^{-20}$ & 9.00$\times$10$^{-20}$ & 1.38$\times$10$^{-18}$ & 1.50$\times$10$^{-17}$\\
0.18 & 8.02$\times$10$^{-17}$ & 1.00$\times$10$^{-16}$ & 1.28$\times$10$^{-16}$ & 8.05$\times$10$^{-19}$ & 1.54$\times$10$^{-18}$ & 2.84$\times$10$^{-18}$ & 1.30$\times$10$^{-18}$ & 2.96$\times$10$^{-17}$ & 3.28$\times$10$^{-16}$\\
0.2 & 2.00$\times$10$^{-15}$ & 2.28$\times$10$^{-15}$ & 2.64$\times$10$^{-15}$ & 3.41$\times$10$^{-17}$ & 5.43$\times$10$^{-17}$ & 9.60$\times$10$^{-17}$ & 2.20$\times$10$^{-17}$ & 6.04$\times$10$^{-16}$ & 6.65$\times$10$^{-15}$\\
0.25 & 8.68$\times$10$^{-13}$ & 9.29$\times$10$^{-13}$ & 1.01$\times$10$^{-12}$ & 5.88$\times$10$^{-14}$ & 7.56$\times$10$^{-14}$ & 1.00$\times$10$^{-13}$ & 3.40$\times$10$^{-14}$ & 3.12$\times$10$^{-13}$ & 3.01$\times$10$^{-12}$\\
0.3 & 5.08$\times$10$^{-11}$ & 5.62$\times$10$^{-11}$ & 6.57$\times$10$^{-11}$ & 9.32$\times$10$^{-12}$ & 1.13$\times$10$^{-11}$ & 1.38$\times$10$^{-11}$ & 5.90$\times$10$^{-12}$ & 2.56$\times$10$^{-11}$ & 2.03$\times$10$^{-10}$\\
0.35 & 9.64$\times$10$^{-10}$ & 1.13$\times$10$^{-09}$ & 1.43$\times$10$^{-09}$ & 3.46$\times$10$^{-10}$ & 4.08$\times$10$^{-10}$ & 4.86$\times$10$^{-10}$ & 2.30$\times$10$^{-10}$ & 6.58$\times$10$^{-10}$ & 4.23$\times$10$^{-09}$\\
0.4 & 9.31$\times$10$^{-09}$ & 1.15$\times$10$^{-08}$ & 1.52$\times$10$^{-08}$ & 5.11$\times$10$^{-09}$ & 5.95$\times$10$^{-09}$ & 6.98$\times$10$^{-09}$ & 3.49$\times$10$^{-09}$ & 7.89$\times$10$^{-09}$ & 4.21$\times$10$^{-08}$\\
0.45 & 5.72$\times$10$^{-08}$ & 7.30$\times$10$^{-08}$ & 9.86$\times$10$^{-08}$ & 4.09$\times$10$^{-08}$ & 4.72$\times$10$^{-08}$ & 5.50$\times$10$^{-08}$ & 2.84$\times$10$^{-08}$ & 5.56$\times$10$^{-08}$ & 2.54$\times$10$^{-07}$\\
0.5 & 2.53$\times$10$^{-07}$ & 3.28$\times$10$^{-07}$ & 4.46$\times$10$^{-07}$ & 2.13$\times$10$^{-07}$ & 2.44$\times$10$^{-07}$ & 2.82$\times$10$^{-07}$ & 1.49$\times$10$^{-07}$ & 2.67$\times$10$^{-07}$ & 1.08$\times$10$^{-06}$\\
0.6 & 2.47$\times$10$^{-06}$ & 3.22$\times$10$^{-06}$ & 4.33$\times$10$^{-06}$ & 2.47$\times$10$^{-06}$ & 2.79$\times$10$^{-06}$ & 3.20$\times$10$^{-06}$ & 1.74$\times$10$^{-06}$ & 2.80$\times$10$^{-06}$ & 9.49$\times$10$^{-06}$\\
0.7 & 1.31$\times$10$^{-05}$ & 1.67$\times$10$^{-05}$ & 2.20$\times$10$^{-05}$ & 1.39$\times$10$^{-05}$ & 1.57$\times$10$^{-05}$ & 1.78$\times$10$^{-05}$ & 9.90$\times$10$^{-06}$ & 1.49$\times$10$^{-05}$ & 4.48$\times$10$^{-05}$\\
0.8 & 4.75$\times$10$^{-05}$ & 5.94$\times$10$^{-05}$ & 7.55$\times$10$^{-05}$ & 5.15$\times$10$^{-05}$ & 5.77$\times$10$^{-05}$ & 6.51$\times$10$^{-05}$ & 3.69$\times$10$^{-05}$ & 5.30$\times$10$^{-05}$ & 1.44$\times$10$^{-04}$\\
0.9 & 1.38$\times$10$^{-04}$ & 1.69$\times$10$^{-04}$ & 2.08$\times$10$^{-04}$ & 1.48$\times$10$^{-04}$ & 1.66$\times$10$^{-04}$ & 1.88$\times$10$^{-04}$ & 1.08$\times$10$^{-04}$ & 1.49$\times$10$^{-04}$ & 3.65$\times$10$^{-04}$\\
1 & 3.46$\times$10$^{-04}$ & 4.19$\times$10$^{-04}$ & 5.10$\times$10$^{-04}$ & 3.65$\times$10$^{-04}$ & 4.11$\times$10$^{-04}$ & 4.73$\times$10$^{-04}$ & 2.73$\times$10$^{-04}$ & 3.63$\times$10$^{-04}$ & 7.95$\times$10$^{-04}$\\
1.25 & 2.33$\times$10$^{-03}$ & 2.86$\times$10$^{-03}$ & 3.70$\times$10$^{-03}$ & 2.33$\times$10$^{-03}$ & 2.77$\times$10$^{-03}$ & 3.43$\times$10$^{-03}$ & 1.81$\times$10$^{-03}$ & 2.41$\times$10$^{-03}$ & 4.02$\times$10$^{-03}$\\
1.5 & 1.04$\times$10$^{-02}$ & 1.33$\times$10$^{-02}$ & 1.78$\times$10$^{-02}$ & (1.45$\times$10$^{-02}$) & (1.79$\times$10$^{-02}$) & (2.21$\times$10$^{-02}$) & 1.17$\times$10$^{-02}$ & 1.57$\times$10$^{-02}$ & 2.64$\times$10$^{-02}$\\
2 & (1.93$\times$10$^{-01}$) & (2.48$\times$10$^{-01}$) & (3.32$\times$10$^{-01}$) & (3.00$\times$10$^{-01}$) & (3.70$\times$10$^{-01}$) & (4.58$\times$10$^{-01}$) & 2.11$\times$10$^{-01}$ & 2.90$\times$10$^{-01}$ & 5.01$\times$10$^{-01}$\\
2.5 & (1.60$\times$10$^{00}$) & (2.06$\times$10$^{00}$) & (2.75$\times$10$^{00}$) & (2.55$\times$10$^{+00}$) & (3.15$\times$10$^{+00}$) & (3.89$\times$10$^{+00}$) & 1.66$\times$10$^{+00}$ & 2.33$\times$10$^{+00}$ & 4.12$\times$10$^{+00}$\\
3 & (7.99$\times$10$^{00}$) & (1.03$\times$10$^{01}$) & (1.37$\times$10$^{01}$) & (1.24$\times$10$^{+01}$) & (1.53$\times$10$^{+01}$) & (1.89$\times$10$^{+01}$ & 7.40$\times$10$^{+00}$ & 1.07$\times$10$^{+01}$ & 1.94$\times$10$^{+01}$\\
3.5 & (2.80$\times$10$^{01}$) & (3.59$\times$10$^{01}$) & (4.81$\times$10$^{01}$) & (4.18$\times$10$^{+01}$) & (5.17$\times$10${+01}$) & 6.39$\times$10$^{+01}$) & 2.34$\times$10$^{+01}$ & 3.44$\times$10$^{+01}$ & 6.42$\times$10$^{+01}$\\
4 & (7.61$\times$10$^{01}$) & (9.75$\times$10$^{01}$) & (1.30$\times$10$^{02}$) & (1.10$\times$10$^{+02}$) & (1.36$\times$10$^{+02}$) & (1.68$\times$10$^{+02}$) & 5.83$\times$10$^{+01}$ & 8.84$\times$10$^{+01}$ & 1.69$\times$10$^{+02}$\\
5 & (3.29$\times$10$^{02}$) & (4.22$\times$10$^{02}$) & (5.65$\times$10$^{02}$) & (4.71$\times$10$^{+02}$) & (5.82$\times$10$^{+02}$) & (7.19$\times$10$^{+02}$) & 2.29$\times$10$^{+02}$ & 3.69$\times$10$^{+02}$ & 7.46$\times$10$^{+02}$\\
6 & (8.95$\times$10$^{02}$) & (1.15$\times$10$^{03}$) & (1.53$\times$10$^{03}$) & (1.33$\times$10$^{+03}$) & (1.64$\times$10$^{+03}$) & (2.03$\times$10$^{+03}$) & 5.90$\times$10$^{+02}$ & 1.02$\times$10$^{+03}$ & 2.19$\times$10$^{+03}$\\
7 & (1.84$\times$10$^{03}$) & (2.36$\times$10$^{03}$) & (3.15$\times$10$^{03}$) & (2.91$\times$10$^{+03}$) & (3.59$\times$10$^{+03}$) & (4.44$\times$10$^{+03}$) & 1.14$\times$10$^{+03}$ & 2.17$\times$10$^{+03}$ & 4.95$\times$10$^{+03}$\\
8 & (3.20$\times$10$^{03}$) & (4.11$\times$10$^{03}$) & (5.49$\times$10$^{03}$) & (5.35$\times$10$^{+03}$) & (6.62$\times$10$^{+03}$) & (8.18$\times$10$^{+03}$) & 1.78$\times$10$^{+03}$ & 3.83$\times$10$^{+03}$ & 9.32$\times$10$^{+03}$\\
9 & (5.13$\times$10$^{03}$) & (6.58$\times$10$^{03}$) & (8.80$\times$10$^{03}$) & (8.68$\times$10$^{+03}$) & (1.07$\times$10$^{+04}$) & (1.33$\times$10$^{+04}$) & 2.36$\times$10$^{+03}$ & 5.92$\times$10$^{+03}$ & 1.55$\times$10$^{+04}$\\
10 & (8.00$\times$10$^{03}$) & (1.03$\times$10$^{04}$) & (1.37$\times$10$^{04}$) & (1.30$\times$10$^{+04}$) & 1.60$\times$10$^{+04}$) & (1.98$\times$10$^{+04}$) & 2.67$\times$10$^{+03}$ & 8.31$\times$10$^{+03}$ & 2.34$\times$10$^{+04}$\\
\end{tabular}
\end{ruledtabular}
\end{table*}
\endgroup

\begingroup
\squeezetable
\begin{table*}
\caption{Monte Carlo reaction rates for the $^{22}$Ne($\alpha$,n)$^{25}$Mg reaction calculated using the Nucleosynthesis Simulator called Starlib ~\citep{starlib}. The median rate represents the recommended ($\alpha$,n) rate determined using the contributions from the 553 keV ($E_x$ = 11167 keV) and the 702 keV ($E_x$ = 11317 keV) resonances observed in the present work along with the other known resonances reported in the literature from the direct measurement of the  $^{22}$Ne($\alpha$,n)$^{25}$Mg reaction. For the 553 keV resonance, the recommended rate was calculated using Jaeger et al.'s upper limit for $\omega\gamma_{(\alpha,n)}$ = 60 neV. The rate values in parenthesis represent the temperatures (T$_9$ $\textgreater$ T$_9$ = 1.5) for which Non-Smoker Hauser Feshbach rates from the JINA Reaclib Database ~\citep{Jin10}, normalized to the experimental results, have been adopted. The Longland et al. ~\citep{PhysRevC.85.065809} and NACRE ~\citep{Angulo19993} rates have also been provided to facilitate the comparison. }
\label{tab:rate_an}
\begin{ruledtabular}
\begin{tabular}{c c c c c c c c c c}
& \multicolumn{3}{c}{Present Work} & \multicolumn{3}{c}{Longland et al. ~\citep{PhysRevC.85.065809}} & \multicolumn{3}{c}{NACRE ~\citep{Angulo19993}}\\
$T_9$ & Low rate  & Median rate & High rate & Low rate  & Median rate & High rate & Low rate  & Median rate & High rate\\  
\hline\\
0.01 & 6.97$\times$10$^{-252}$ & 7.87$\times$10$^{-252}$ & 8.86$\times$10$^{-252}$ & 0.00 & 0.00 & 0.00 & 0.00 & 0.00 & 0.00\\
0.011 & 6.65$\times$10$^{-230}$ & 7.50$\times$10$^{-230}$ & 8.45$\times$10$^{-230}$ & 0.00 & 0.00 & 0.00 & 0.00 & 0.00 & 0.00\\
0.012 & 1.39$\times$10$^{-211}$ & 1.57$\times$10$^{-211}$ & 1.77$\times$10$^{-211}$ & 0.00 & 0.00 & 0.00 & 0.00 & 0.00 & 0.00\\
0.013 & 4.34$\times$10$^{-196}$ & 4.90$\times$10$^{-196}$ & 5.51$\times$10$^{-196}$ & 0.00 & 0.00 & 0.00 & 0.00 & 0.00 & 0.00\\
0.014 & 8.47$\times$10$^{-183}$ & 9.56$\times$10$^{-183}$ & 1.08$\times$10$^{-182}$ & 0.00 & 0.00 & 0.00 & 0.00 & 0.00 & 0.00\\
0.015 & 2.81$\times$10$^{-171}$ & 3.17$\times$10$^{-171}$ & 3.57$\times$10$^{-171}$ & 0.00 & 0.00 & 0.00 & 0.00 & 0.00 & 0.00\\
0.016 & 3.40$\times$10$^{-161}$ & 3.84$\times$10$^{-161}$ & 4.32$\times$10$^{-161}$ & 0.00 & 0.00 & 0.00 & 0.00 & 0.00 & 0.00\\
0.018 & 2.20$\times$10$^{-144}$ & 2.48$\times$10$^{-144}$ & 2.79$\times$10$^{-144}$ & 0.00 & 0.00 & 0.00 & 0.00 & 0.00 & 0.00\\
0.02 & 6.26$\times$10$^{-131}$ & 7.07$\times$10$^{-131}$ & 7.96$\times$10$^{-131}$ & 0.00 & 0.00 & 0.00 & 0.00 & 0.00 & 0.00\\
0.025 & 1.09$\times$10$^{-106}$ & 1.23$\times$10$^{-106}$ & 1.38$\times$10$^{-106}$ & 0.00 & 0.00 & 0.00 & 0.00 & 0.00 & 0.00\\
0.03 & 1.66$\times$10$^{-90}$ & 1.88$\times$10$^{-90}$ & 2.11$\times$10$^{-90}$ & 5.12$\times$10$^{-88}$ & 5.08$\times$10$^{-87}$ & 2.25$\times$10$^{-86}$ & 0.00 & 0.00 & 0.00\\
0.04 & 3.50$\times$10$^{-70}$ & 3.99$\times$10$^{-70}$ & 4.90$\times$10$^{-70}$ & 1.46$\times$10$^{-67}$ & 1.49$\times$10$^{-66}$ & 6.64$\times$10$^{-66}$ & 0.00 & 0.00 & 0.00\\
0.05 & 2.62$\times$10$^{-57}$ & 3.17$\times$10$^{-57}$ & 6.03$\times$10$^{-57}$ & 2.99$\times$10$^{-55}$ & 3.05$\times$10$^{-54}$ & 1.36$\times$10$^{-53}$ & 0.00 & 0.00 & 0.00\\
0.06 & 3.16$\times$10$^{-48}$ & 3.88$\times$10$^{-48}$ & 7.98$\times$10$^{-48}$ & 4.92$\times$10$^{-47}$ & 4.87$\times$10$^{-46}$ & 2.17$\times$10$^{-45}$ & 0.00 & 0.00 & 0.00\\
0.07 & 1.05$\times$10$^{-41}$ & 1.30$\times$10$^{-41}$ & 2.69$\times$10$^{-41}$ & 3.70$\times$10$^{-41}$ & 3.48$\times$10$^{-40}$ & 1.55$\times$10$^{-39}$ & 0.00 & 0.00 & 0.00\\
0.08 & 8.09$\times$10$^{-37}$ & 9.97$\times$10$^{-37}$ & 2.07$\times$10$^{-36}$ & 1.03$\times$10$^{-36}$ & 8.44$\times$10$^{-36}$ & 3.73$\times$10$^{-35}$ & 0.00 & 0.00 & 0.00\\
0.09 & 5.00$\times$10$^{-33}$ & 6.17$\times$10$^{-33}$ & 1.28$\times$10$^{-32}$ & 3.23$\times$10$^{-33}$ & 2.19$\times$10$^{-32}$ & 9.43$\times$10$^{-32}$ & 0.00 & 0.00 & 0.00\\
0.1 & 5.32$\times$10$^{-30}$ & 6.56$\times$10$^{-30}$ & 1.36$\times$10$^{-29}$ & 2.17$\times$10$^{-30}$ & 1.20$\times$10$^{-29}$ & 4.92$\times$10$^{-29}$ & 0.00 & 0.00 & 0.00\\
0.11 & 1.57$\times$10$^{-27}$ & 1.94$\times$10$^{-27}$ & 4.02$\times$10$^{-27}$ & 4.65$\times$10$^{-28}$ & 2.12$\times$10$^{-27}$ & 8.22$\times$10$^{-27}$ & 0.00 & 0.00 & 0.00\\
0.12 & 1.78$\times$10$^{-25}$ & 2.19$\times$10$^{-25}$ & 4.56$\times$10$^{-25}$ & 4.24$\times$10$^{-26}$ & 1.62$\times$10$^{-25}$ & 5.82$\times$10$^{-25}$ & 1.00$\times$10$^{-26}$ & 2.33$\times$10$^{-25}$ & 1.20$\times$10$^{-22}$\\
0.13 & 9.61$\times$10$^{-24}$ & 1.19$\times$10$^{-23}$ & 2.48$\times$10$^{-23}$ & 1.94$\times$10$^{-24}$ & 6.61$\times$10$^{-24}$ & 2.14$\times$10$^{-23}$ & 4.00$\times$10$^{-25}$ & 8.64$\times$10$^{-24}$ & 5.37$\times$10$^{-21}$\\
0.14 & 2.92$\times$10$^{-22}$ & 3.63$\times$10$^{-22}$ & 7.60$\times$10$^{-22}$ & 5.27$\times$10$^{-23}$ & 1.64$\times$10$^{-22}$ & 4.81$\times$10$^{-22}$ & 1.10$\times$10$^{-23}$ & 1.96$\times$10$^{-22}$ & 1.44$\times$10$^{-19}$\\
0.15 & 5.61$\times$10$^{-21}$ & 7.00$\times$10$^{-21}$ & 1.48$\times$10$^{-20}$ & 9.94$\times$10$^{-22}$ & 2.74$\times$10$^{21}$ & 7.18$\times$10$^{-21}$ & 2.60$\times$10$^{-22}$ & 3.03$\times$10$^{-21}$ & 2.55$\times$10$^{-18}$\\
0.16 & 7.42$\times$10$^{-20}$ & 9.38$\times$10$^{-20}$ & 2.01$\times$10$^{-19}$ & 1.43$\times$10$^{-20}$ & 3.39$\times$10$^{-20}$ & 7.89$\times$10$^{-20}$ & 4.80$\times$10$^{-21}$ & 3.51$\times$10$^{-20}$ & 3.20$\times$10$^{-17}$\\
0.18 & 5.54$\times$10$^{-18}$ & 7.35$\times$10$^{-18}$ & 1.63$\times$10$^{-17}$ & 1.61$\times$10$^{-18}$ & 2.74$\times$10$^{-18}$ & 5.01$\times$10$^{-18}$ & 9.10$\times$10$^{-19}$ & 2.68$\times$10$^{-18}$ & 2.22$\times$10$^{-15}$\\
0.2 & 1.82$\times$10$^{-16}$ & 2.62$\times$10$^{-16}$ & 6.01$\times$10$^{-16}$ & 9.14$\times$10$^{-17}$ & 1.24$\times$10$^{-16}$ & 1.79$\times$10$^{-16}$ & 7.00$\times$10$^{-17}$ & 1.23$\times$10$^{-16}$ & 6.70$\times$10$^{-14}$\\
0.25 & 1.41$\times$10$^{-13}$ & 2.49$\times$10$^{-13}$ & 5.66$\times$10$^{-13}$ & 1.68$\times$10$^{-13}$ & 2.06$\times$10$^{-13}$ & 2.53$\times$10$^{-13}$ & 1.82$\times$10$^{-13}$ & 2.30$\times$10$^{-13}$ & 3.14$\times$10$^{-11}$\\
0.3 & 1.82$\times$10$^{-11}$ & 3.33$\times$10$^{-11}$ & 6.84$\times$10$^{-11}$ & 2.74$\times$10$^{-11}$ & 3.36$\times$10$^{-11}$ & 4.15$\times$10$^{-11}$ & 3.37$\times$10$^{-11}$ & 4.06$\times$10$^{-11}$ & 1.92$\times$10$^{-09}$\\
0.35 & 6.93$\times$10$^{-10}$ & 1.20$\times$10$^{-09}$ & 2.21$\times$10$^{-09}$ & 1.05$\times$10$^{-09}$ & 1.29$\times$10$^{-09}$ & 1.59$\times$10$^{-09}$ & 1.37$\times$10$^{-09}$ & 1.64$\times$10$^{-09}$ & 3.68$\times$10$^{-08}$\\
0.4 & 1.12$\times$10$^{-08}$ & 1.81$\times$10$^{-08}$ & 3.05$\times$10$^{-08}$ & 1.64$\times$10$^{-08}$ & 2.00$\times$10$^{-08}$ & 2.45$\times$10$^{-08}$ & 2.18$\times$10$^{-08}$ & 2.60$\times$10$^{-08}$ & 3.43$\times$10$^{-07}$\\
0.45 & 1.04$\times$10$^{-07}$ & 1.54$\times$10$^{-07}$ & 2.39$\times$10$^{-07}$ & 1.42$\times$10$^{-07}$ & 1.71$\times$10$^{-07}$ & 2.07$\times$10$^{-07}$ & 1.90$\times$10$^{-07}$ & 2.27$\times$10$^{-07}$ & 1.98$\times$10$^{-06}$\\
0.5 & 6.74$\times$10$^{-07}$ & 9.16$\times$10$^{-07}$ & 1.30$\times$10$^{-06}$ & 8.51$\times$10$^{-07}$ & 1.00$\times$10$^{-06}$ & 1.19$\times$10$^{-06}$ & 1.12$\times$10$^{-06}$ & 1.33$\times$10$^{-06}$ & 8.26$\times$10$^{-06}$\\
0.6 & 1.72$\times$10$^{-05}$ & 1.97$\times$10$^{-05}$ & 2.33$\times$10$^{-05}$ & 1.74$\times$10$^{-05}$ & 1.92$\times$10$^{-05}$ & 2.15$\times$10$^{-05}$ & 2.10$\times$10$^{-05}$ & 2.45$\times$10$^{05}$ & 7.97$\times$10$^{-05}$\\
0.7 & 2.79$\times$10$^{-04}$ & 2.94$\times$10$^{-04}$ & 3.12$\times$10$^{-04}$ & 2.36$\times$10$^{-04}$ & 2.51$\times$10$^{-04}$ & 2.69$\times$10$^{-04}$ & 2.67$\times$10$^{-04}$ & 3.04$\times$10$^{-04}$ & 5.60$\times$10$^{-04}$\\
0.8 & 2.76$\times$10$^{-03}$ & 2.85$\times$10$^{-03}$ & 2.95$\times$10$^{-03}$ & 2.15$\times$10$^{-03}$ & 2.27$\times$10$^{-03}$ & 2.42$\times$10$^{-03}$ & 2.39$\times$10$^{-03}$ & 2.69$\times$10$^{-03}$ & 3.63$\times$10$^{-03}$\\
0.9 & 1.79$\times$10$^{-02}$ & 1.85$\times$10$^{-02}$ & 1.91$\times$10$^{-02}$ & 1.36$\times$10$^{-02}$ & 1.43$\times$10$^{-02}$ & 1.51$\times$10$^{-02}$ & 1.50$\times$10$^{-02}$ & 1.68$\times$10$^{-02}$ & 2.00$\times$10$^{-02}$\\
1 & 8.36$\times$10$^{-02}$ & 8.68$\times$10$^{-02}$ & 9.00$\times$10$^{-02}$ & 6.34$\times$10$^{-02}$ & 6.64$\times$10$^{-02}$ & 6.98$\times$10$^{-02}$ & 6.99$\times$10$^{-02}$ & 7.81$\times$10$^{-02}$ & 8.91$\times$10$^{-02}$\\
1.25 & 1.51$\times$10$^{00}$ & 1.59$\times$10$^{00}$ & 1.68$\times$10$^{00}$ & 1.18$\times$10$^{+00}$ & 1.22$\times$10$^{+00}$ & 1.27$\times$10$^{+00}$ & 1.33$\times$10$^{+00}$ & 1.50$\times$10$^{+00}$ & 1.68$\times$10$^{+00}$\\
1.5 & 1.14$\times$10$^{01}$ & 1.22$\times$10$^{01}$ & 1.30$\times$10$^{01}$ & (1.09$\times$10$^{+01}$) & (1.14$\times$10$^{+01}$) & (1.18$\times$10$^{+01}$) & 1.12$\times$10$^{+01}$ & 1.30$\times$10$^{+01}$ & 1.48$\times$10$^{+01}$\\
2 & (3.17$\times$10$^{02}$) & (3.39$\times$10$^{02}$) & (3.63$\times$10$^{02}$) & (2.92$\times$10$^{+02}$) & (3.04$\times$10$^{+02}$) & (3.16$\times$10$^{+02}$) & 2.22$\times$10$^{+02}$ & 2.76$\times$10$^{+02}$ & 3.30$\times$10$^{+02}$ \\
2.5 & (3.15$\times$10$^{03}$) & (3.37$\times$10$^{03}$) & (3.60$\times$10$^{03}$) & (2.74$\times$10$^{+03}$) & (2.85$\times$10$^{+03}$) & (2.96$\times$10$^{+03}$) & 2.03$\times$10$^{+03}$ & 2.55$\times$10$^{+03}$ & 3.07$\times$10$^{+03}$\\
3 & (1.70$\times$10$^{04}$) & (1.82$\times$10$^{04}$) & (1.95$\times$10$^{04}$) & (1.41$\times$10$^{+04}$) & (1.46$\times$10$^{+04}$) & (1.52$\times$10$^{+04}$) & 1.01$\times$10$^{+04}$ & 1.28$\times$10$^{+04}$ & 1.55$\times$10$^{+04}$\\
3.5 & (6.20$\times$10$^{04}$) & (6.63$\times$10$^{04}$) & (7.10$\times$10$^{04}$) & (4.96$\times$10$^{+04}$) & (5.16$\times$10$^{+04}$) & (5.37$\times$10$^{+04}$) & 3.46$\times$10$^{+04}$ & 4.44$\times$10$^{+04}$ & 5.42$\times$10$^{+04}$\\
4 & (1.72$\times$10$^{05}$) & (1.85$\times$10$^{05}$) & (1.98$\times$10$^{05}$) & (1.36$\times$10$^{+05}$) & (1.41$\times$10$^{+05}$) & (1.47$\times$10$^{+05}$) & 9.40$\times$10$^{+04}$ & 1.22$\times$10$^{+05}$ & 1.50$\times$10$^{+05}$\\
5 & (7.87$\times$10$^{05}$) & (8.42$\times$10$^{05}$) & (9.02$\times$10$^{05}$) & (6.10$\times$10$^{+05}$) & (6.34$\times$10$^{+05}$) & (6.59$\times$10$^{+05}$) & 4.30$\times$10$^{+05}$ & 5.70$\times$10$^{+05}$ & 7.11$\times$10$^{+05}$\\
6 & (2.32$\times$10$^{06}$) & (2.48$\times$10$^{06}$) & (2.66$\times$10$^{06}$) & (1.80$\times$10$^{+06}$) & (1.88$\times$10$^{+06}$) & (1.95$\times$10$^{+06}$) & 1.28$\times$10$^{+06}$ & 1.74$\times$10$^{+06}$ & 2.20$\times$10$^{+06}$\\
7 & (5.29$\times$10$^{06}$) & (5.66$\times$10$^{06}$) & (6.06$\times$10$^{06}$) & (4.07$\times$10$^{+06}$) & (4.23$\times$10$^{+06}$) & (4.40$\times$10$^{+06}$) & 2.88$\times$10$^{+06}$ & 4.02$\times$10$^{+06}$ & 5.16$\times$10$^{+06}$\\
8 & (1.03$\times$10$^{07}$) & (1.11$\times$10$^{07}$) & (1.18$\times$10$^{07}$) & (7.70$\times$10$^{+06}$) & (8.01$\times$10$^{+06}$) & (8.32$\times$10$^{+06}$) & 5.37$\times$10$^{+06}$ & 7.69$\times$10$^{+06}$ & 1.00$\times$10$^{+07}$\\
9 & (1.84$\times$10$^{07}$) & (1.97$\times$10$^{07}$) & (2.11$\times$10$^{07}$) & (1.28$\times$10$^{+07}$) & (1.33$\times$10$^{+07}$) & (1.39$\times$10$^{+07}$) & 8.80$\times$10$^{+06}$ & 1.29$\times$10$^{+07}$ & 1.70$\times$10$^{+07}$\\
10 & (3.11$\times$10$^{07}$) & (3.33$\times$10$^{07}$) & (3.56$\times$10$^{07}$) & (1.97$\times$10$^{+07}$) & (2.04$\times$10$^{+07}$) & (2.12$\times$10$^{+07}$) & 1.29$\times$10$^{+07}$ & 1.96$\times$10$^{+07}$ & 2.63$\times$10$^{+07}$\\
\end{tabular}
\end{ruledtabular}
\end{table*}
\endgroup

\begin{figure}
\includegraphics[width= 1\columnwidth]{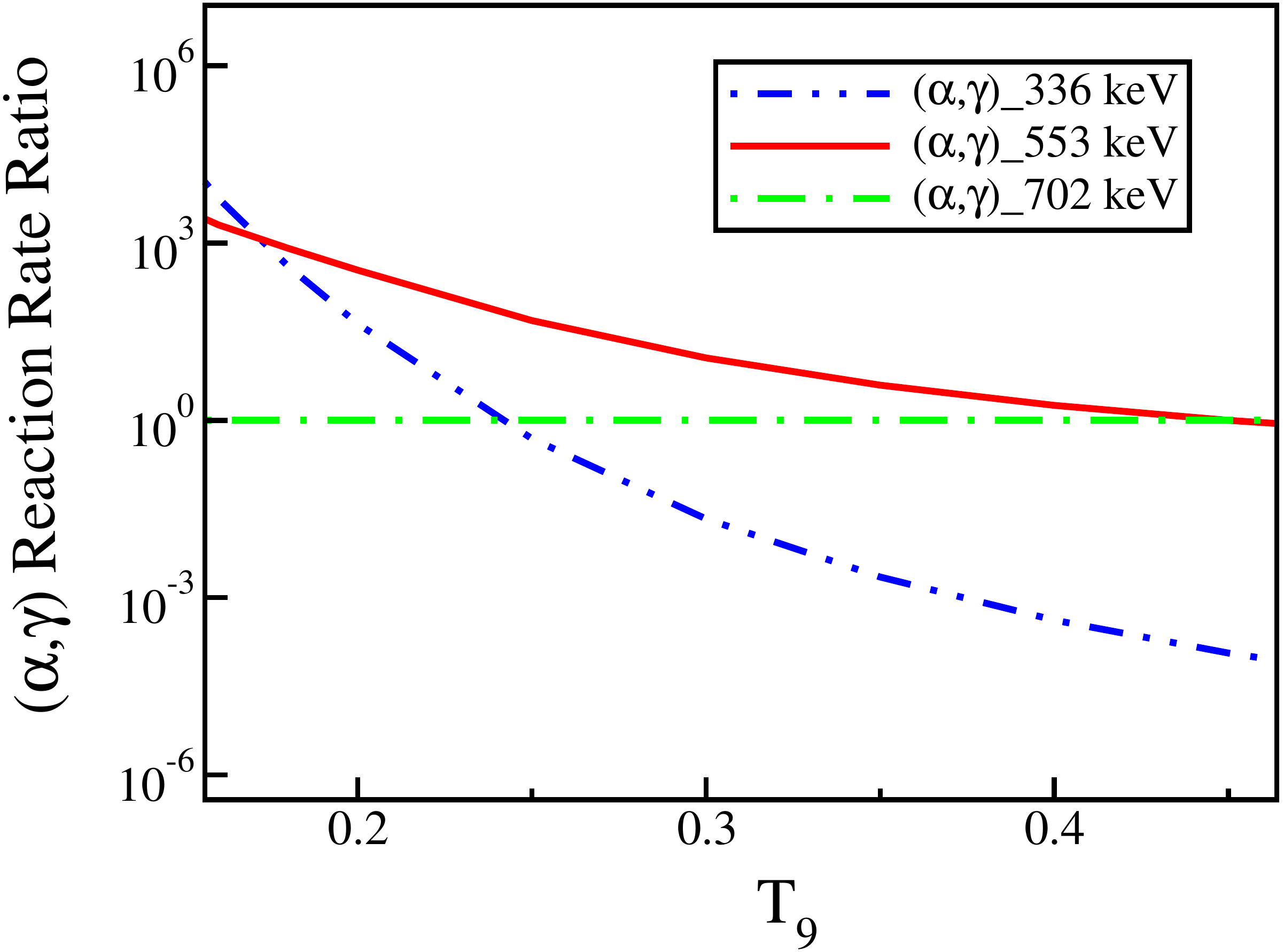}
    \caption{(Color online) Comparison between the reaction rates corresponding to individual resonances observed in the present work, above the $\alpha$-threshold, normalized to the $E_{\alpha}$ = 703 keV resonance which is the lowest directly observed resonance.}
    \label{fig:ag_individual_plot}
 \end{figure}
 
\begin{figure}
\includegraphics[width= 1\columnwidth]{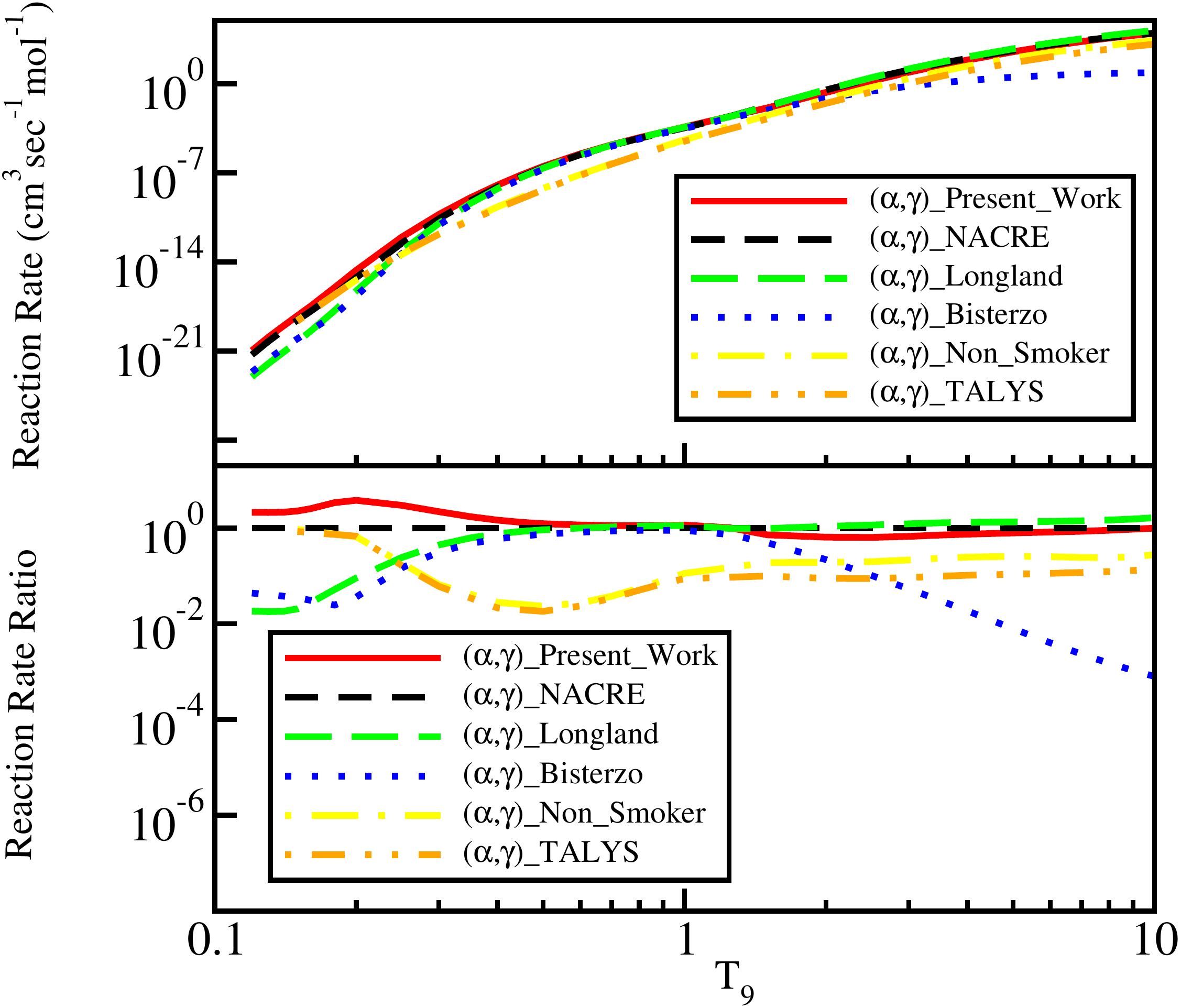}
    \caption{(Color online) The upper panel shows the comparison of the total $^{22}$Ne($\alpha,\gamma$)$^{26}$Mg reaction rate calculated for the present work with the rates available in the literature. The lower panel shows the same comparison normalized to the NACRE total ($\alpha,\gamma$) rate ~\citep{Angulo19993}.}
    \label{fig:ag_plot}
 \end{figure} 

\begin{figure}
\includegraphics[width= 1\columnwidth]{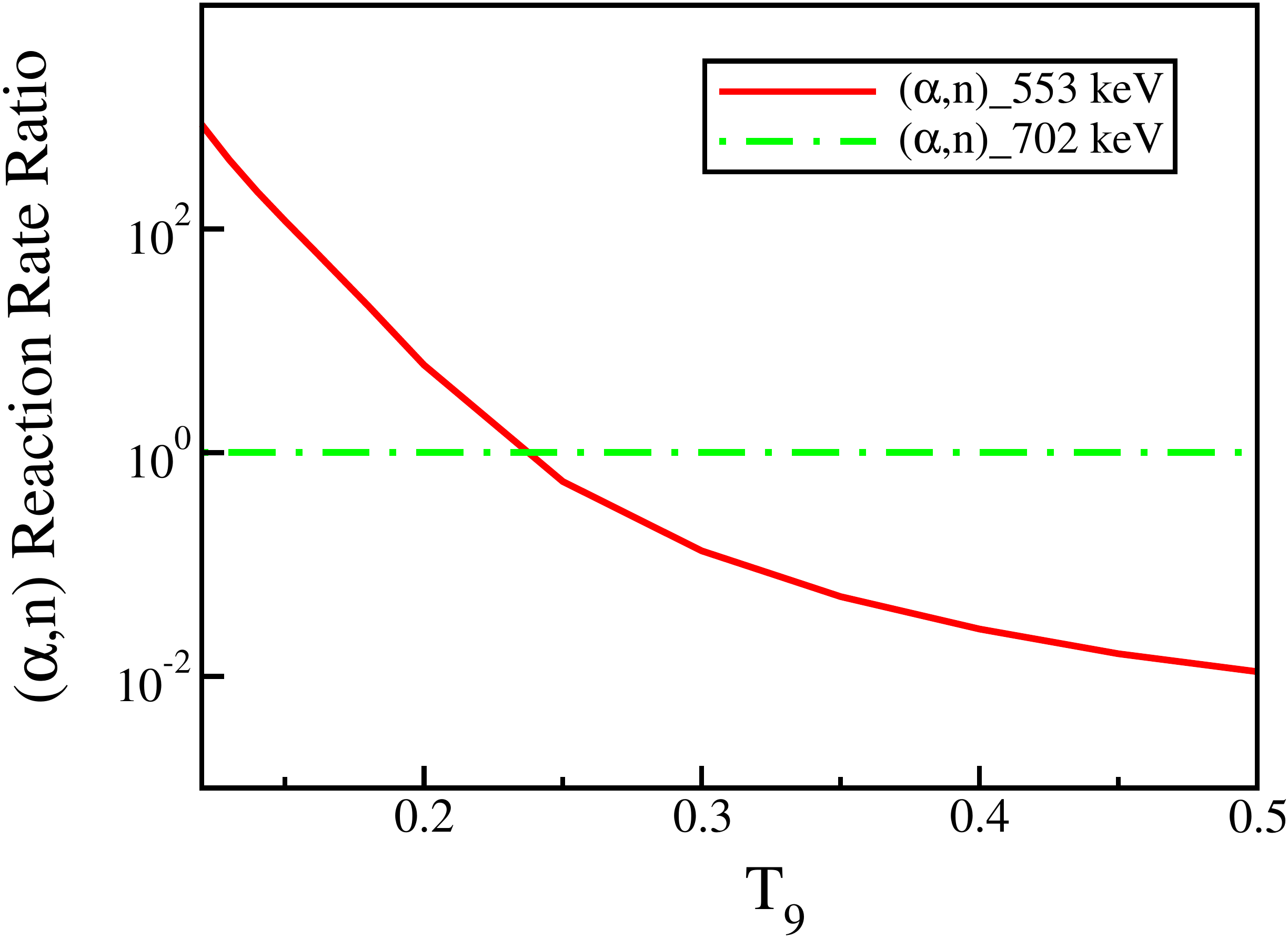}
    \caption{(Color online) Comparison between the reaction rates corresponding to individual resonances observed in the present work, above the $n$-threshold, normalized to the $E_{\alpha}$ = 703 keV resonance which is the lowest directly observed resonance. }
    \label{fig:an_individual_plot}
 \end{figure}

\begin{figure}
\includegraphics[width= 1\columnwidth]{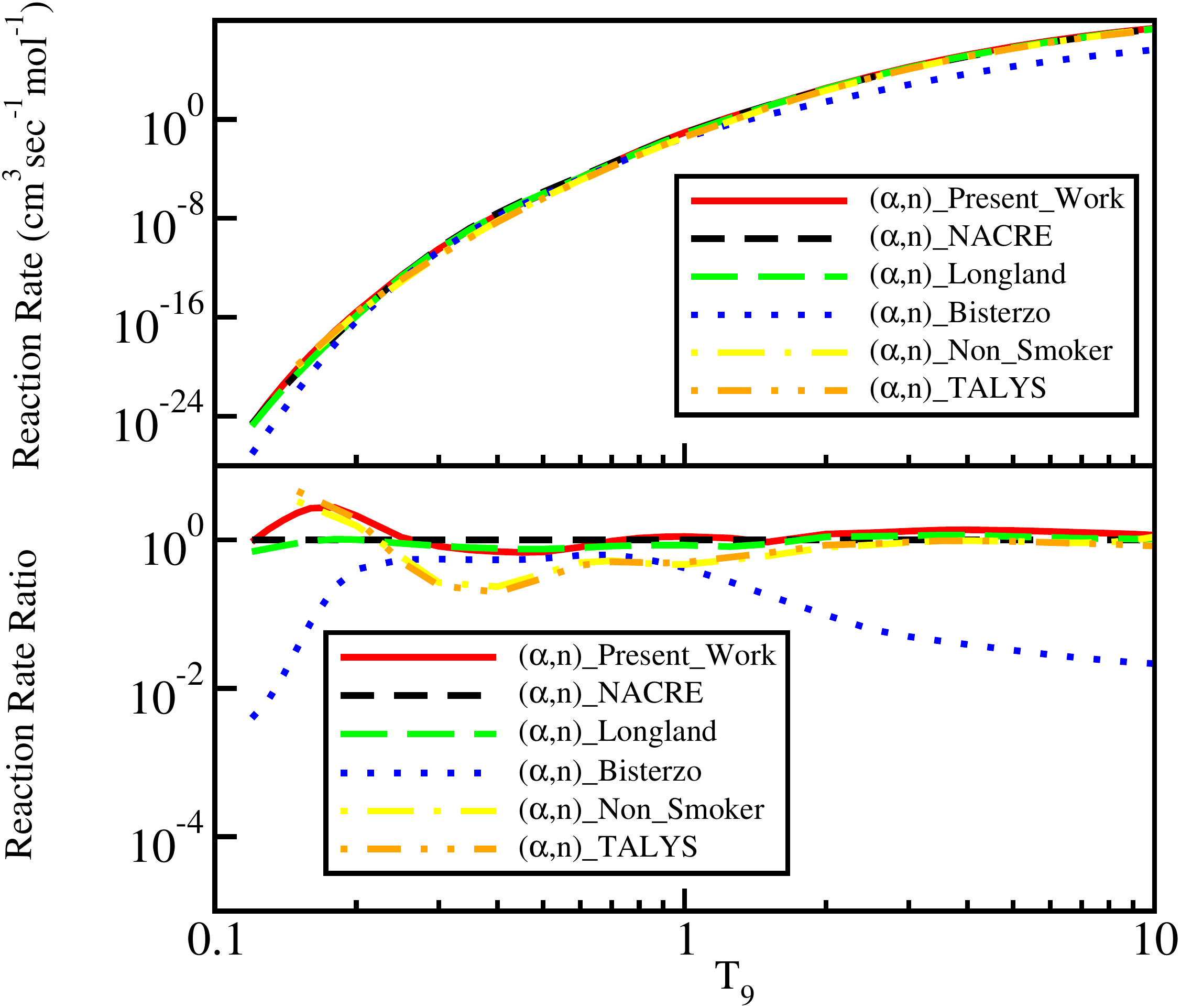}
    \caption{(Color online) The upper panel shows the comparison of the total $^{22}$Ne($\alpha$,n)$^{25}$Mg reaction rate calculated for the present work with the rates available in the literature. The lower panel shows the same comparison normalized to the NACRE total ($\alpha$,n) rate ~\citep{Angulo19993}.}
    \label{fig:an_plot}
 \end{figure}  
 
 \begin{figure}
\includegraphics[width= 1\columnwidth]{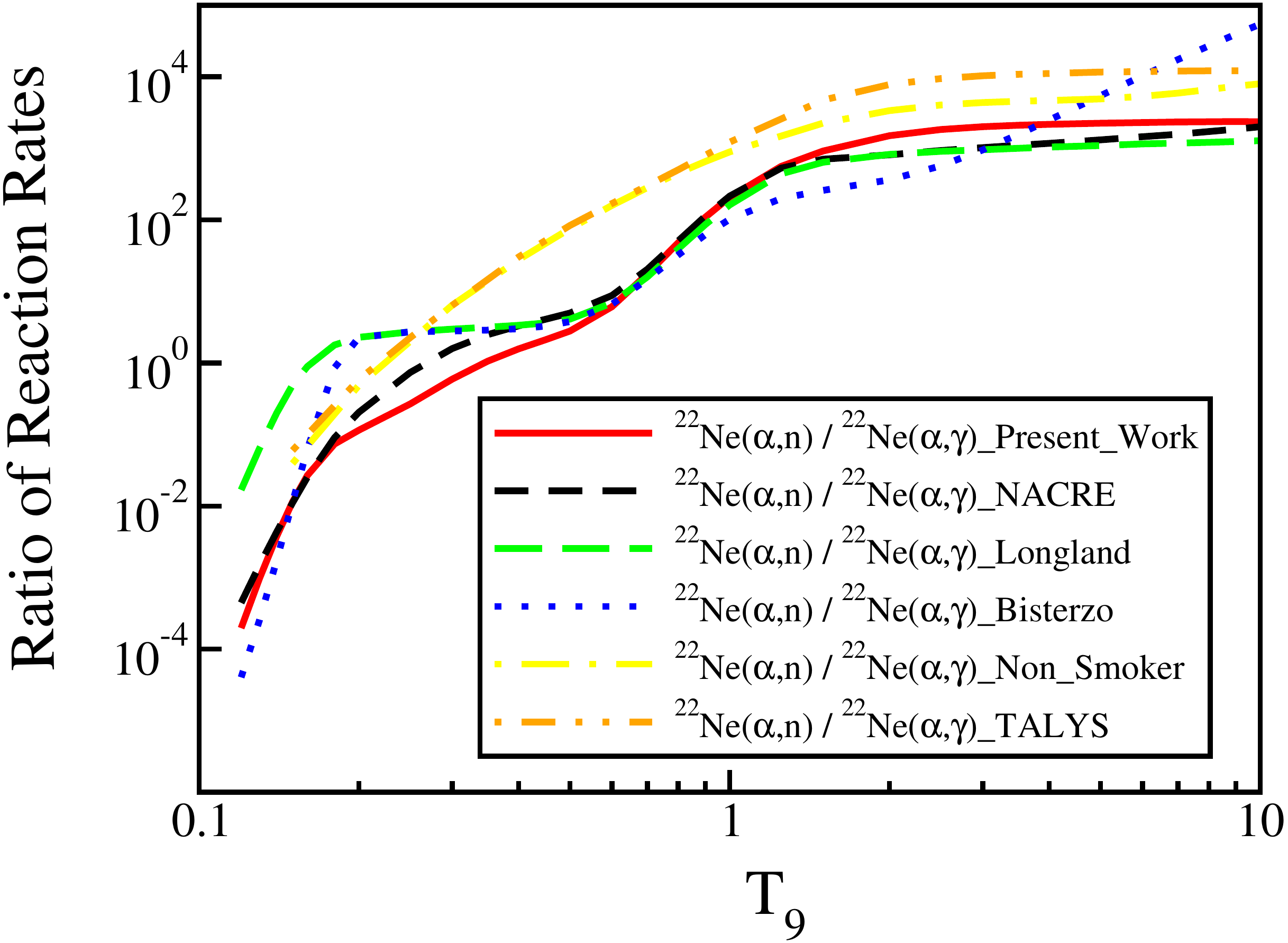}
    \caption{(Color online) Comparison of the $^{22}$Ne($\alpha$,n)/$^{22}$Ne($\alpha,\gamma$) reaction rate ratio for the present work with the literature rate ratios.}
    \label{fig:rate_plot}
 \end{figure} 

Fig. ~\ref{fig:ag_individual_plot} depicts the behaviour of the present $^{22}$Ne($\alpha,\gamma)^{26}$Mg reaction rate with respect to temperature, corresponding to individual resonances observed in the astrophysical region of interest. Each rate has been normalized to the ($\alpha,\gamma$) rate corresponding to the $E_{\alpha}$ = 703 keV resonance wich is the lowest directly observed resonance. For $T_9$ $\textless$ 0.18, the $E_R$ = 336 keV resonance ($E_x$ = 10951 keV) (blue dash double dot line) has the largest contribution to the ($\alpha,\gamma$) rate. However, for 0.18 $\textless$ $T_9$ $\textless$ 0.4, the reaction rate corresponding to the $E_R$ = 553 keV resonance ($E_x$ = 11167 keV) (red line) dominates. The effect of this can also be seen in Fig. ~\ref{fig:ag_plot}. The upper panel in Fig. ~\ref{fig:ag_plot} shows the comparison of the total $^{22}$Ne($\alpha,\gamma$)$^{26}$Mg reaction rate calculated for the present work (red solid line) with that calculated by Longland et al. ~\citep{PhysRevC.85.065809} (green long dash line), Bisterzo et al. ~\citep{Bisterzo} (blue dotted line) and NACRE ~\citep{Angulo19993} (black small dash line) along with the Hauser Feshbach rates (Non-Smoker from JINA REACLIB ~\citep{Jin10} (yellow dash dot line) and Talys ~\citep{Tal08} (orange dash double dot line). The lower panel shows the same comparison normalized to the NACRE total ($\alpha,\gamma$) rate ~\citep{Angulo19993}, to facilitate the comparison. Unlike Bisterzo et al. data, all the other rates have adopted the Hauser Feshbach rates, normalized to their respective experimental data, for temperatures approximately above 1.25 GK. The present total ($\alpha,\gamma$) rate is higher, by almost 2 orders of magnitude, than the Longland et al. and Bisterzo et al. rates and almost by a factor of 3 than the NACRE rates. This is due to the large $\alpha$-width associated with the $E_R$ = 553 keV resonance (as can be seen in Table ~\ref{tab:widths}). 

Fig. ~\ref{fig:an_individual_plot} depicts the behaviour of the present $^{22}$Ne($\alpha$,n)$^{25}$Mg reaction rate with respect to temperature, corresponding to individual resonances observed above the $n$-threshold. Each rate has been normalized to the ($\alpha$,n) rate corresponding to the $E_{\alpha}$ = 703 KeV resonance wich is the lowest directly observed resonance. For $T_9$ $\textless$ 0.22, the reaction rate corresponding to the $E_R$ = 553 keV resonance ($E_x$ = 11167 keV) dominates above which the rate corresponding to the $E_R$ = 702 keV resonance ($E_x$ = 11317 keV) dominates. The same trend can be seen in Fig. ~\ref{fig:an_plot} showing a similar comparison between the ($\alpha$,n) rates in the upper and lower panels as shown in Fig. ~\ref{fig:ag_plot} for the ($\alpha,\gamma$) rates. 

The behaviour of the $^{22}$Ne($\alpha$,n)/$^{22}$Ne($\alpha,\gamma$) reaction rate ratios is shown in Fig. ~\ref{fig:rate_plot}. For $T_9$ $\textless$ 0.5, the ($\alpha,n$)/($\alpha,\gamma$) rate ratio in the present work (red solid line) is lower than that corresponding to  Longland et al. ~\citep{PhysRevC.85.065809} (green long dash line), Bisterzo et al. ~\citep{Bisterzo} (blue dotted line) and NACRE ~\citep{Angulo19993} (black small dash line)  rate ratios. This is because the $\alpha$-width of the $E_R$ = 553 keV resonance ($E_x$ = 11167 keV)  significantly increases the ($\alpha,\gamma$) rate such that for $T_9 \textless$ 0.35 the $^{22}$Ne($\alpha,\gamma$) rate dominates over the $^{22}$Ne($\alpha$,n) rate. This strongly indicates the need to study the influence of low energy resonances near the $\alpha$-threshold on the $\alpha$-capture rates, which has been the primary objective of the present measurements.\\

\subsection{Astrophysical Implications}

As mentioned in section \ref{sec:intro}, $^{22}$Ne($\alpha$,n)$^{26}$Mg is believed to be the main neutron source in massive stars and AGB stars of intermediate mass. In low mass AGB stars with solar like metallicities, it is marginally activated during the advanced thermal pulses giving rise to a small neutron exposure with a high peak neutron density (N$_n$(peak) $\sim$ 10$^{10}$ cm$^{-3}$. As a consequence, the impact of $^{22}$Ne+$\alpha$ capture rates on the whole s-process distribution is marginal in these models, with the exception of a few isotopes involved in the branches of the s-path. In the following paragraphs, a comparison of the affect of the present $^{22}$Ne+$\alpha$ capture rates and literature rates on the s-process nucleosynthesis in these astrophysical scenarios has been presented.

\begin{figure}
\includegraphics[width= 1\columnwidth]{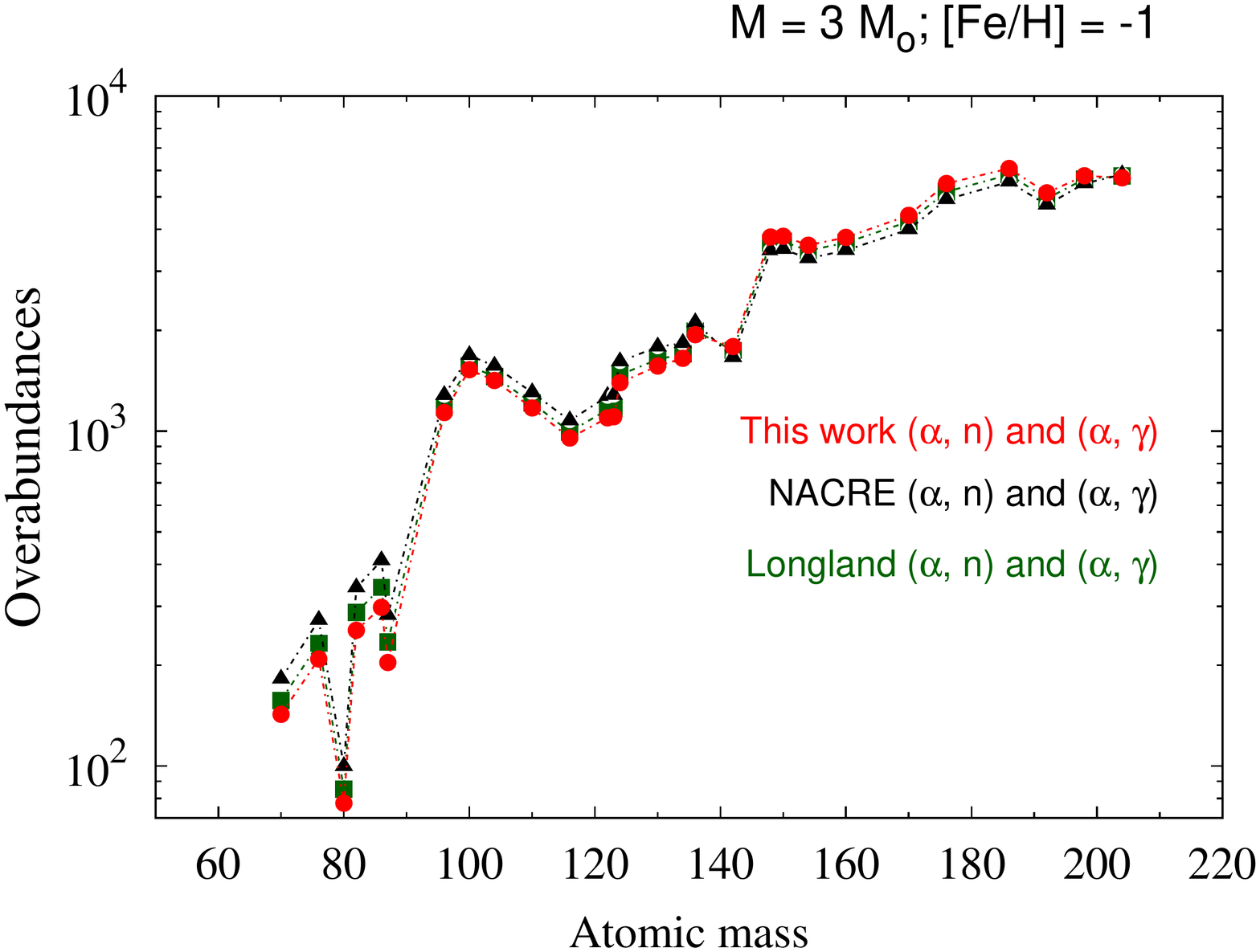}
    \caption{(Color online) Impact of $^{22}$Ne+$\alpha$ capture rates on the isotopic over-abundances for a 3 M$_{\odot}$ AGB star at 1/10 solar metallicity.  Comparison is shown between the impacts due to present $\alpha$-capture rates,  Longland et al. rates ~\citep{PhysRevC.85.065809} and NACRE rates ~\citep{Angulo19993}. }
    \label{fig:agb_3}
 \end{figure}

\begin{figure}
\includegraphics[width= 1\columnwidth]{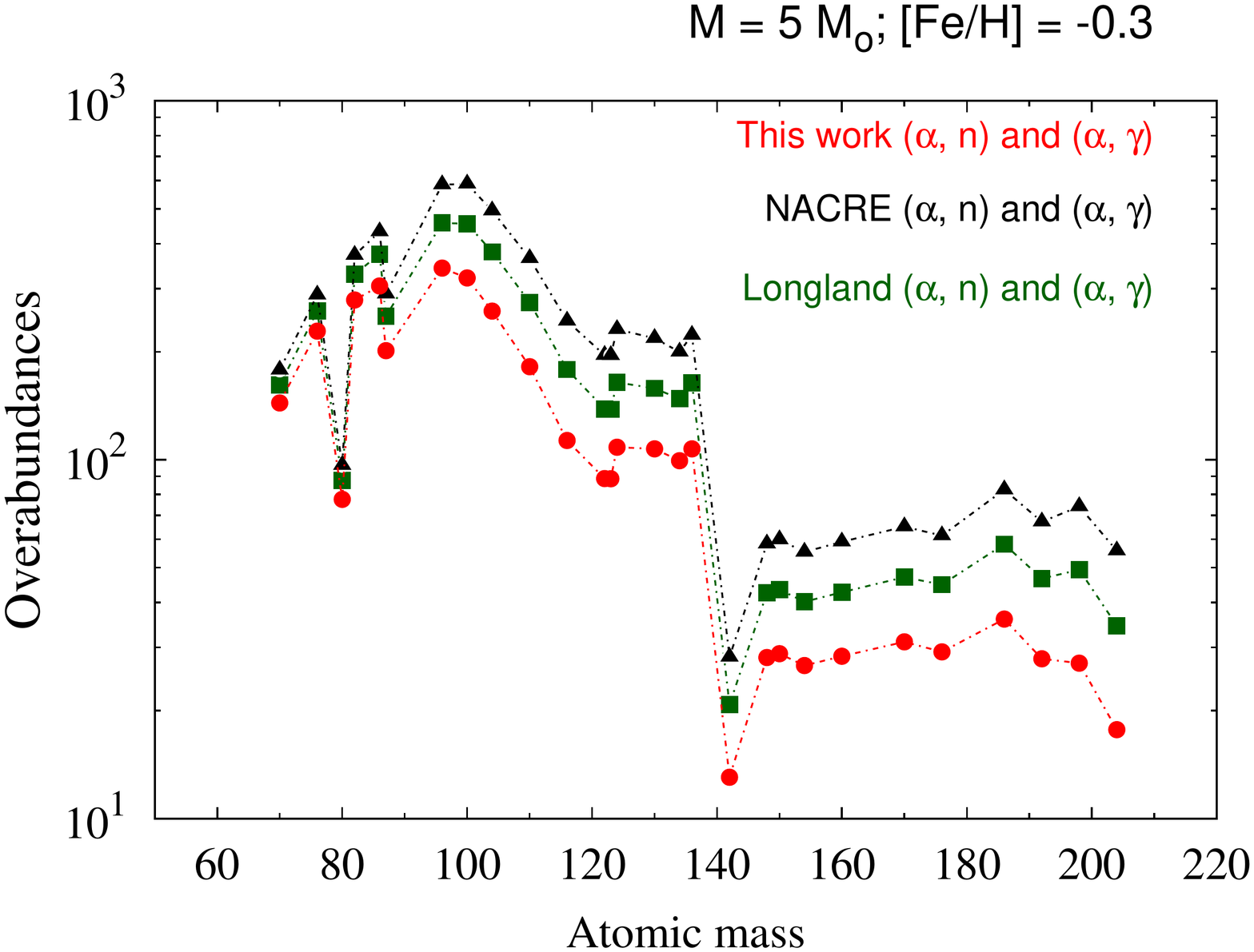}
    \caption{(Color online) Impact of $^{22}$Ne+$\alpha$ capture rates on the isotopic over-abundances for a 5 M$_{\odot}$ AGB star at half solar metalliicity. Comparison is shown between the impacts due to present $\alpha$-capture rates,  Longland et al. rates ~\citep{PhysRevC.85.065809} and NACRE rates ~\citep{Angulo19993}.}
    \label{fig:agb_5}
 \end{figure}

\begin{figure}
\includegraphics[width= 1\columnwidth]{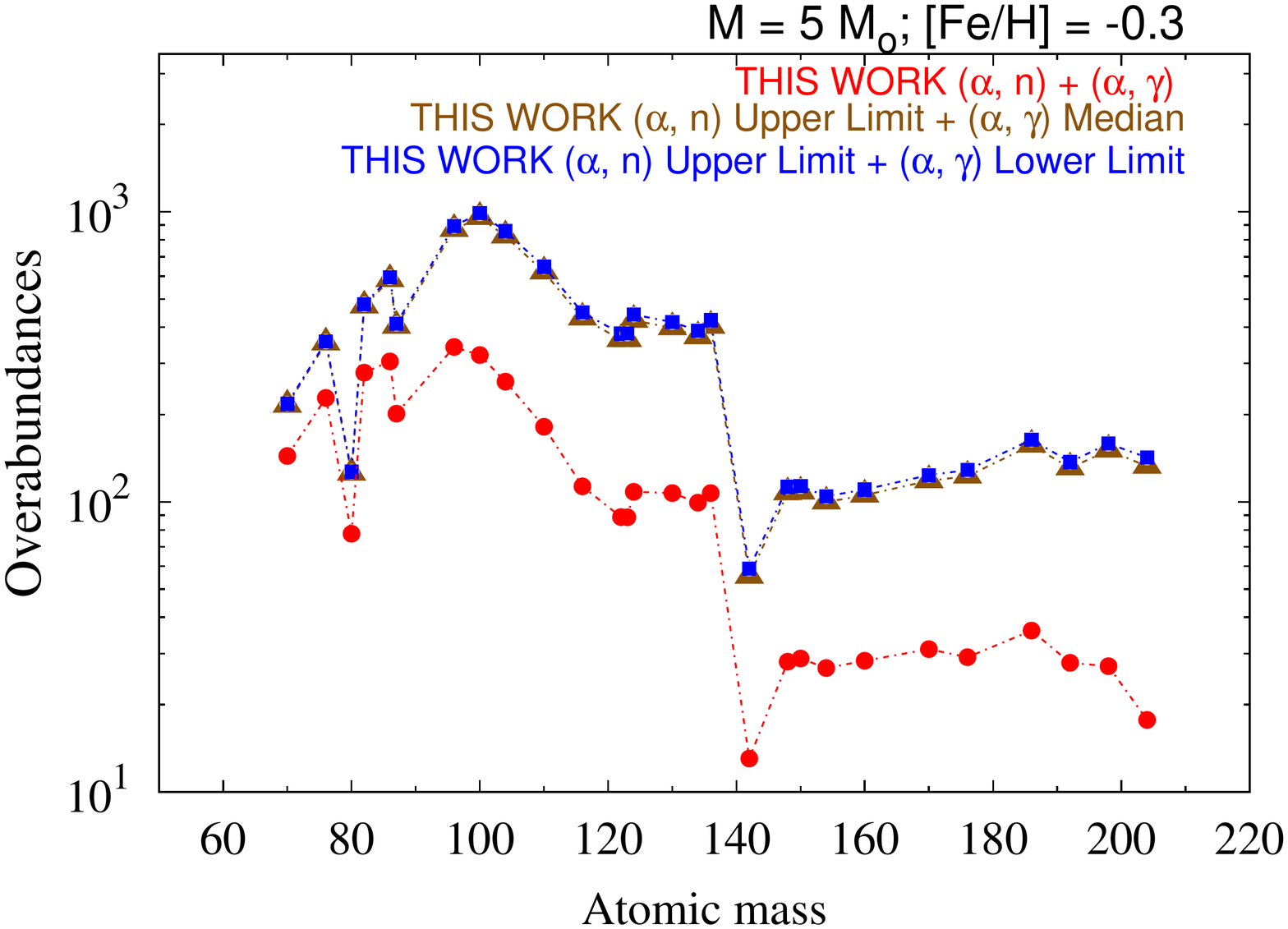}
    \caption{(Color online) Impact of present ($\alpha$,n) upper limit and ($\alpha,\gamma$) lower limit on the isotopic over-abundances for a 5 M$_{\odot}$ AGB star at half solar metallicity. }
    \label{fig:agb_5_high}
 \end{figure}

\begin{figure}
\includegraphics[width= 1\columnwidth]{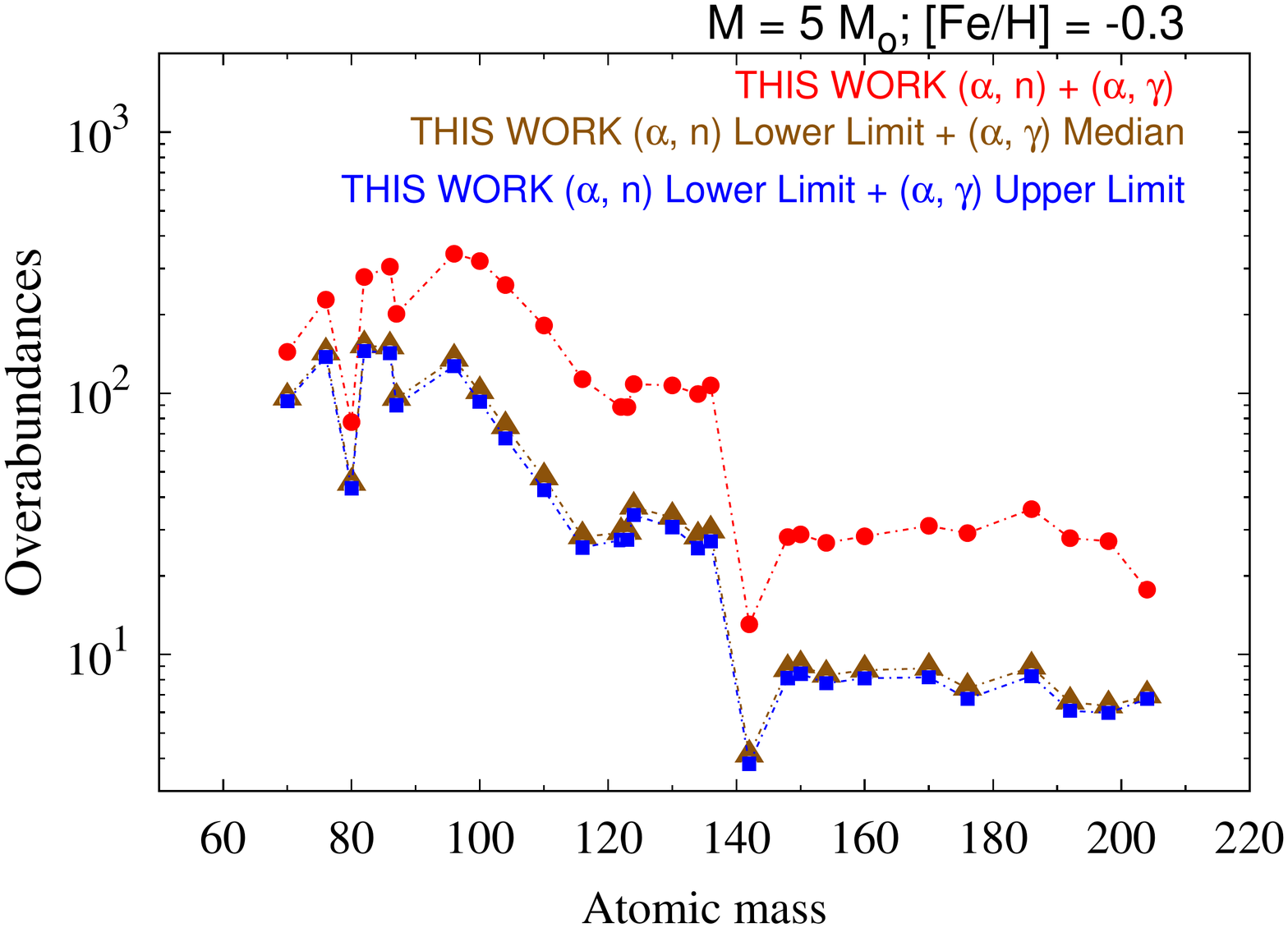}
    \caption{(Color online) Impact of present ($\alpha$,n) lower limit and ($\alpha,\gamma$) upper limit on the isotopic over-abundances for a 5 M$_{\odot}$ AGB star at half solar metallicity.}
    \label{fig:agb_5_low}
 \end{figure}

Figures ~\ref{fig:agb_3}, ~\ref{fig:agb_5}, ~\ref{fig:agb_5_high} and ~\ref{fig:agb_5_low} depict the impact of $^{22}$Ne+$\alpha$ capture rates on isotopic over-abundances of low and intermediate mass AGB stars. The over-abundances signify the mass fractions (X$_i$) over the solar-scaled initial values. 
 
For a given AGB initial mass, the maximum temperature at the bottom of the convective zone increases as the metallicity decreases, and the $^{22}$Ne($\alpha$,n)$^{26}$Mg source becomes more efficient. For a 3 M$_{\odot}$ AGB model at [Fe/H] = -1, the maximum temperature at the bottom of the advanced thermal pulses reaches $T_9$ $\sim$ 0.35. In the present scenario, both $^{22}$Ne($\alpha$,n)$^{26}$Mg and $^{13}$C($\alpha$,n)$^{16}$O neutron sources compete. The resulting variations in the over-abundances are nevertheless small, as can be seen in Fig. ~\ref{fig:agb_3}, because the contribution of $^{13}$C($\alpha$,n)$^{16}$O dominates.

For a 5 M$_{\odot}$ AGB model at [Fe/H] = -0.3, higher temperatures are readily achieved at the bottom of the thermal pulses ($T_9$ $\approx$ 0.35). As a result the $^{22}$Ne($\alpha$,n)$^{25}$Mg reaction is efficiently activated producing higher peak neutron densities of $\approx$ 10$^{11}$ cm$^{-3}$. However, as can be seen in Fig. ~\ref{fig:ag_plot}, the present recommended ($\alpha,\gamma$) rate is larger than that recommended by NACRE and Longland et al. Hence, it strongly competes with the ($\alpha$,n) neutron source leading to a decrease in the over-abundances corresponding to the present work, as shown in Fig. ~\ref{fig:agb_5}.

 \begin{figure}
\includegraphics[width= 1\columnwidth]{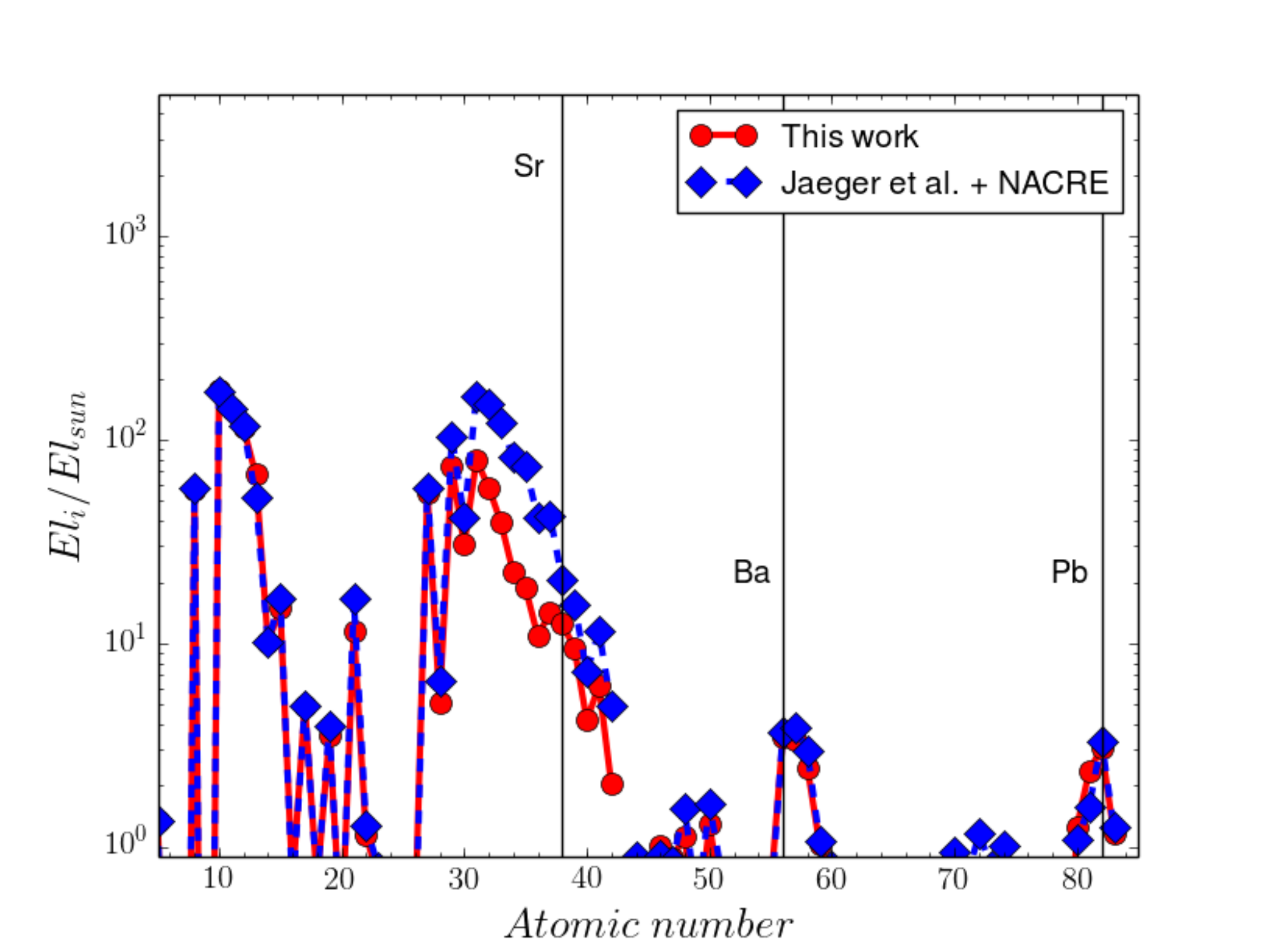}
    \caption{(Color online) Impact of $^{22}$Ne+$\alpha$ capture rates on the s-process distribution for a 25 M$_{\odot}$ massive star. Comparison is shown between the impacts due to the present $\alpha$-capture rates and a combination of Jaeger et al. $^{22}$Ne($\alpha$,n) rates ~\citep{PhysRevLett.87.202501} and $^{22}$Ne($\alpha,\gamma$) NACRE rates ~\citep{Angulo19993}. }
    \label{fig:marco_nacre}
 \end{figure}
 
  \begin{figure}
\includegraphics[width= 1\columnwidth]{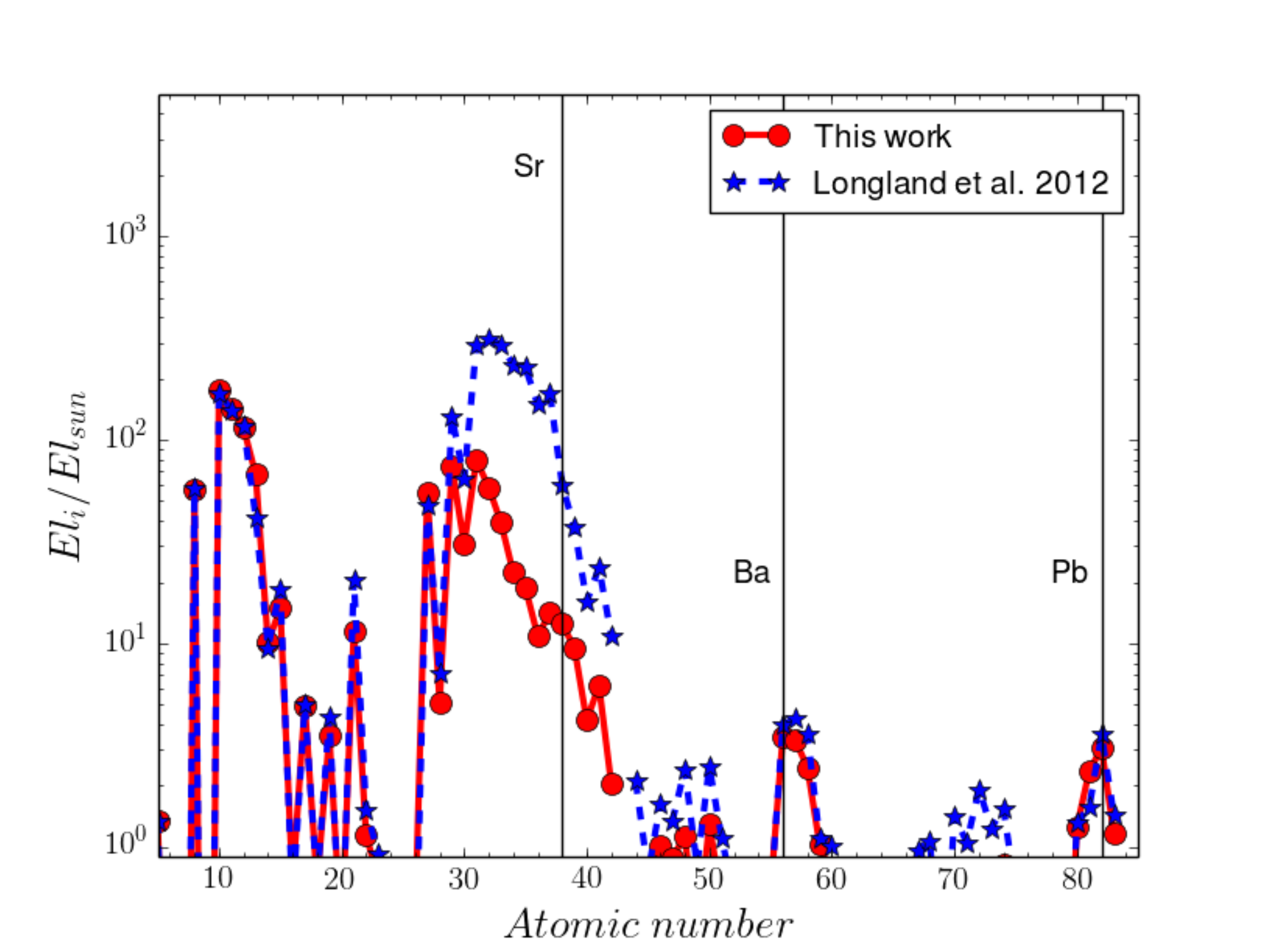}
    \caption{(Color online) Impact of $^{22}$Ne+$\alpha$ capture rates on the s-process distribution for a 25 M$_{\odot}$ massive star. Comparison is shown between the impacts due to present $\alpha$-capture rates and Longland et al. rates ~\citep{PhysRevC.85.065809}. }
    \label{fig:marco_longland}
 \end{figure}
 
  \begin{figure}
\includegraphics[width= 1\columnwidth]{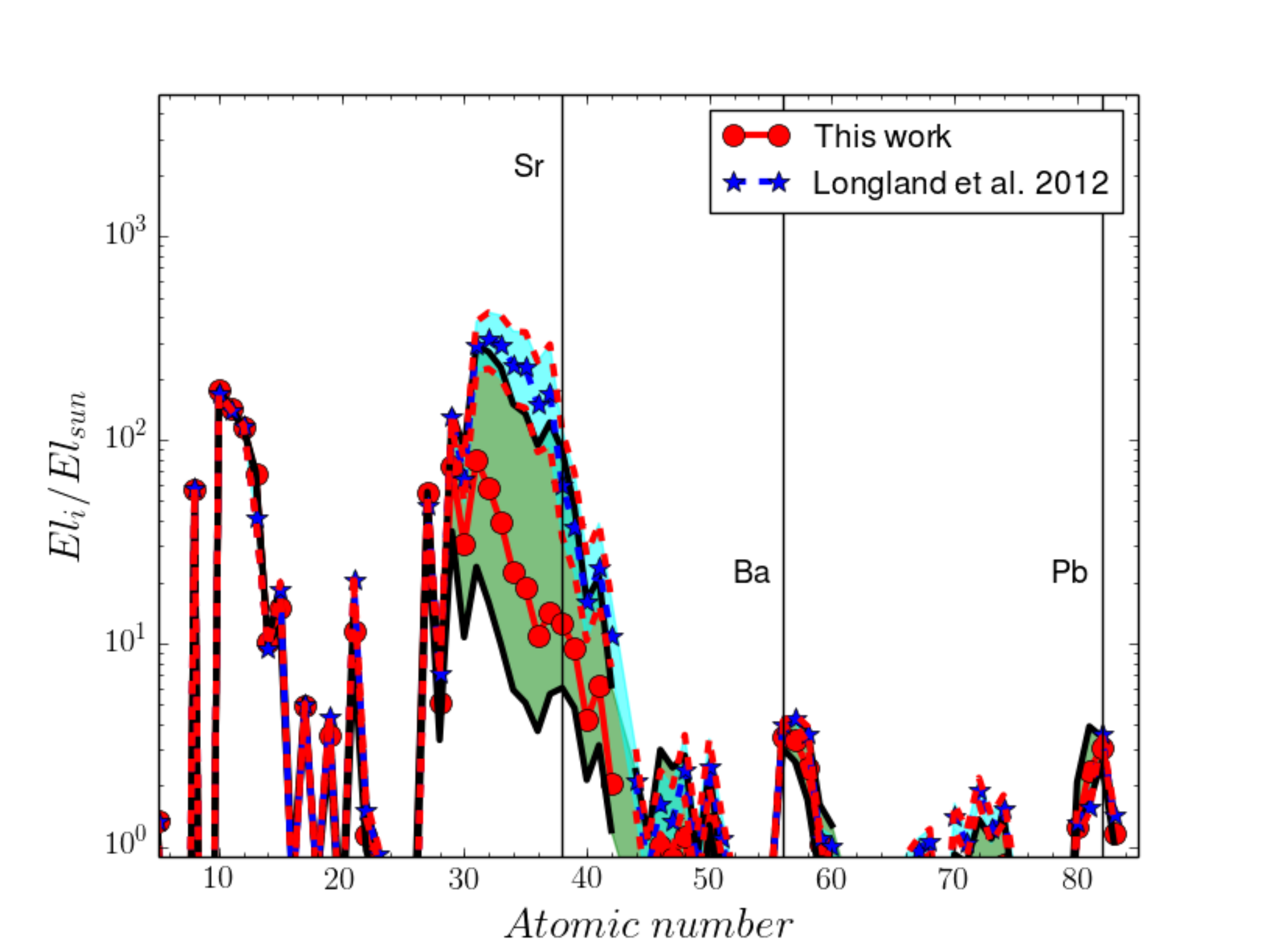}
    \caption{(Color online) Uncertainty range in the s-process distribution corresponding to Longland et al. ~\citep{PhysRevC.85.065809} (light blue region) and present $^{22}$Ne+$\alpha$ capture rates (green region). }
    \label{fig:marco_uncertainity}
 \end{figure}
 
Figures ~\ref{fig:marco_longland}, ~\ref{fig:marco_nacre} and ~\ref{fig:marco_uncertainity} illustrate the impact of $^{22}$Ne+$\alpha$ capture rates on the isotopic over-abundance for a 25 M$_{\odot}$, Z = 0.02 massive star which includes contribution from both the convective core He-burning as well as from the He-core ashes in the convective C-burning shell. However, as has been discussed in section ~\ref{sec:intro}, under C-burning conditions in massive stars, the $^{22}$Ne(p,$\gamma)^{23}$Na reaction becomes the main competitor of the $^{22}$Ne neutron source instead of the $^{22}$Ne($\alpha,\gamma$)$^{26}$Mg reaction. Hence, the present ($\alpha,\gamma$) rates do not change the overall contribution coming from C-burning.  In all these figures, $El_i$ / $El_{sun}$ represents the elemental over-abundance with respect to the solar abundance. The weak s-process region is between Fe and Sr-Y-Zr, where there is high production efficiency. As can be seen in Figures  ~\ref{fig:ag_plot} and ~\ref{fig:rate_plot}, the present recommended ($\alpha,\gamma$) rate at $T_9$ = 0.3 is stronger than the corresponding to Longland et al. and NACRE rates. Hence, it strongly impacts the availability of $^{22}$Ne for the s-process in He-burning conditions, thereby showing a decrease in the over-abundances for the present rates as can be seen in Figures ~\ref{fig:marco_longland} and ~\ref{fig:marco_nacre}.

The low rates and high rates associated with the median (recommended) $^{22}$Ne+$\alpha$ capture rates in Tables ~\ref{tab:rate_ag} and ~\ref{tab:rate_an} translate into the resulting uncertainties associated with the s-process distribution shown in Figures ~\ref{fig:agb_5_high}, ~\ref{fig:agb_5_low} and ~\ref{fig:marco_uncertainity}. Figures ~\ref{fig:agb_5_high} and ~\ref{fig:agb_5_low} illustrate the uncertainty regions in the 5 M$_{\odot}$ AGB star for different combinations of present ($\alpha$,n) lower and upper limits, and present ($\alpha,\gamma$) median rates, and lower and upper limits. Fig. ~\ref{fig:marco_uncertainity} illustrates the uncertainty band for the present work (green area) along with that corresponding to Longland et al. (blue area) in the 25 M$_{\odot}$ massive star for  $^{22}$Ne($\alpha$,n) high - $^{22}$Ne($\alpha,\gamma$) low and $^{22}$Ne($\alpha$,n) low - $^{22}$Ne($\alpha,\gamma$) high range. 

In all these figures, the present $^{22}$Ne($\alpha$,n) + $^{22}$Ne($\alpha,\gamma$) rates strongly favour the reduction of s-process over-abundances associated with massive stars as well as AGB stars of intermediate initial mass. This is due to the large $\alpha$-width associated with $E_x$ = 11167 keV which significantly increases the ($\alpha,\gamma$) rate thereby reducing the efficiency of ($\alpha$,n) rate. The dominant contribution to the present rate uncertainties comes from the discrepancy associated with the $\Gamma_{\gamma}$ and the $\Gamma_n$ values corresponding to $E_x$ = 11167 keV level. Using Jaeger et al. upper limit for $\omega\gamma_{(\alpha,n)}$ = 60 neV ~\citep{PhysRevLett.87.202501} and $\Gamma_{\alpha}$ = 2 $\times$ 10$^{-7}$ (Table ~\ref{tab:widths}), the $\Gamma_{\gamma}$ / $\Gamma_n$ ratio should be approximately equal to 9. On the other hand, the values adopted from reference ~\cite{PhysRevC.85.044615}, as explained in section \ref{subsec:discussion}, gives $\Gamma_{\gamma}$ / $\Gamma_n$ = 2 (2) / 0.6 (0.4) =  3 resulting in $\omega\gamma_{(\alpha,n)}$ = 125 neV. The affect of such disparities is  illustrated by the green area in Fig. ~\ref{fig:marco_uncertainity} which almost coincides with the uncertainty band of Longland et al. (light blue area), and the brown triangles and blue squares in Figures ~\ref{fig:agb_5_low} and ~\ref{fig:agb_5_high}. 

All these results clearly emphasize the need to not only determine the resonances that could have a notable impact on the $^{22}$Ne+$\alpha$ capture rates but also the need to get a better handle on their associated resonance parameters.     

\section{CONCLUSIONS} \label{sec:con}
 
The goal of this study was to investigate the nuclear structure of $^{26}$Mg and determine the $\alpha$-widths for the resonances observed above the $\alpha$-threshold. This nucleus is the compound nucleus that is formed during $\alpha$-capture reactions on $^{22}$Ne that is predicted to serve as the primary neutron source for the s-process in massive stars and intermediate mass AGB stars.

In the present work, six resonances have been observed above the $\alpha$-threshold, with four ($E_x$ = 10717 (9), 10822 (10), 10951 (21) and 11085 (8) keV) between the alpha and neutron thresholds and two ($E_x$ = 11167 (8) and 11317 (18) keV) above the neutron threshold. 

Among the six observed resonances, the $E_x$ = 10951, 11167 and 11317 keV exhibited pronounced $\alpha$-cluster structures, as reflected by their large $\alpha$-spectroscopic factors (Table ~\ref{tab:widths}. Hence, these resonances dominated the $\alpha$-capture rates with the $E_x$ = 11167 keV increasing the ($\alpha,\gamma$) rate by almost 2 orders of magnitude above the Longland et al. ~\citep{PhysRevC.85.065809} and Bisterzo et al. ~\citep{Bisterzo} rates and almost by a factor of 3 above the NACRE rates ~\citep{Angulo19993}. The rate contributions corresponding to the $E_x$ = 10822 and 11085 keV were included in the uncertainty calculations since these resonances were observed only in the ($\alpha,\alpha'$) measurement. 

A similar trend was seen in the s-process elemental distribution. The present $^{22}$Ne($\alpha$,n) + $^{22}$Ne($\alpha,\gamma$) rates favoured reduced s-process over-abundances in massive stars and intermediate mass AGB stars where $T_9$ $\geq$ 0.3 is readily achieved to activate the $^{22}$Ne neutron source. On the other hand, in low mass AGB stars, where such high temperatures are reached only during the last few thermal pulses, the s-process over-abundances corresponding to the present rates did not show much variations compared to the literature rates. 

All in all, the recommended $^{22}$Ne+$\alpha$ capture rates, determined in the present measurements, strongly suggest a reduction in the number of $^{22}$Ne nuclei available for neutron production thereby lowering the s-process over abundances. However, the associated uncertainties point towards the need to better constraint the resonance parameters in order to establish the efficiency of $^{22}$Ne neutron source in a more assertive manner. One of the future efforts being planned in this direction is the proposition to study $^{22}$Ne($\alpha,\gamma$)$^{26}$Mg reaction in inverse kinematics using the 5U accelerator, the helium jet gas target ~\citep{Kontos2012272} and the St. George Recoil Separator ~\citep{Couder200835} developed at the University of Notre Dame. The promising ability of the St. George Separator to effectively separate the beam from the reaction products will help reduce the beam induced background thereby allowing a better study of the $^{22}$Ne+$\alpha$ low energy resonances in the direct reaction channel.

\section*{ACKNOWLEDGEMENTS}
A sincere thanks to the RCNP staff for all their effort in delivering high quality alpha and $^6$Li dispersion matched beams necessary for the present measurements.  This work was funded by the National Science Foundation (NSF) through Grant No. PHY-1068192, and the Joint Institute for Nuclear Astrophysics (JINA) Grant No. PHY-0822648. Marco Pignatari acknowledges the siginificant support to NuGrid from NSF grants : PHY-09-22648 (JINA), PHY-1430152 (JINA Center for the Evolution of the Elements) and EU MIRG-CT-2006-046520. He also acknowledges the support from the "Lend$\ddot{u}$let-2014" Programme of the Hungarian Academy of Sciences (Hungary) and from SNF (Switzerland).  Sara Bisterzo acknowledges the support from PHY-0822648 (JINA) as well as from B2FH Association for the numerical calculations.

\bibliography{references_1}  
\end{document}